\documentclass[onecolumn,superscriptaddress,aps]{revtex4-2}
\usepackage{graphicx}
\usepackage{amsmath}
\usepackage{amsthm}
\usepackage{amssymb}
\usepackage{latexsym}
\usepackage{array}
\usepackage{hyperref}
\usepackage{float}
\usepackage{amsfonts}
\usepackage{dsfont}
\usepackage{mathrsfs}
\usepackage{verbatim}
\usepackage{bbold}
\usepackage{lipsum}
\usepackage[normalem]{ulem}

\usepackage{upgreek}

\usepackage{makecell}
\usepackage{adjustbox,lipsum}

\usepackage{algorithm}
\usepackage{algpseudocode}

\usepackage{color}

\newcommand{\bra}[1]{\left<#1\right|}
\newcommand{\ket}[1]{\left|#1\right>}
\newcommand{\abs}[1]{\left|#1\right|}
\newcommand{\norm}[1]{\left\lVert#1\right\rVert}

\newcommand{\braket}[2]{\left<{#1}|{#2}\right>}
\newcommand{\ketbra}[2]{\ket{#1}\!\!\bra{#2}}

\newtheorem{theorem}{Theorem}
\newtheorem{proposition}{Proposition}
\newtheorem{lemma}{Lemma}
\newtheorem{corollary}{Corollary}
\newtheorem{definition}{Definition}

\newcommand{\tr}[1]{\mbox{Tr}{#1}}

\begin{document}

\title{Quantum Interference Needs Convention: Overlap-Determinability and Unified No-Superposition Principle}

\author{Jeongho~Bang}\email{jbang@yonsei.ac.kr}
\affiliation{Institute for Convergence Research and Education in Advanced Technology, Yonsei University, Seoul 03722, Republic of Korea}
\affiliation{Department of Quantum Information, Yonsei University, Incheon 21983, Republic of Korea}

\author{Kyoungho~Cho}
\affiliation{Institute for Convergence Research and Education in Advanced Technology, Yonsei University, Seoul 03722, Republic of Korea}
\affiliation{Department of Statistics and Data Science, Yonsei University, Seoul 03722, Republic of Korea}

\author{Ki~Hyuk~Yee}\email{quick11@kias.re.kr}
\affiliation{School of Computational Sciences, Korea Institute for Advanced Study, Seoul 02455, Korea}

\date{\today}

\begin{abstract}
Quantum superposition is often phrased as the ability to add state vectors. In practice, however, the physical quantity is a ray (a rank-one projector), so each input specifies only a projector and leaves a gauge freedom in the phases of its vector representatives. This becomes a real operational barrier when one asks for a device that, given two independently prepared unknown pure states, outputs a coherent state proportional to a prescribed linear combination. We identify the missing ingredient as not probabilistic but phase-like. One needs a physical scenario that fixes a single phase convention on the relevant set of rays, so that the overlaps become well defined complex numbers. Thus, we formalize this through phase conventions and a single notion---dubbed as ``overlap-determinability.'' Our main theorem gives an exact equivalence: A nonzero completely positive trace-nonincreasing map that probabilistically produces superposition on a domain exists if and only if that domain is overlap-determinable. This unifies modern no-superposition results and characterizes the exceptional yes-go protocols, which succeed precisely when side information supplies the required missing resource. We then show that granting universal access to such convention-fixed overlaps destabilizes the familiar foundational and computational constraints. It enables forbidden transformations akin to quantum cloning and yields super-luminal signaling. It would also permit reflections about unknown states, leading to exponentially fast overlap amplification and a collapse of Grover's search lower bound to a logarithmic query complexity.
\end{abstract}

\maketitle

\section{Introduction}

Quantum information processing owes much of its distinctive power to the ability to prepare and manipulate the quantum superpositions. In a quantum circuit, a single register can support a vast set of computational branches simultaneously, but the advantage is not mere parallelism: it is the controlled interference between branches---tuned by relative phases and overlaps---that turns a large state space into an observable bias. This viewpoint is particularly transparent in amplitude-amplification-type algorithms, where a tiny overlap is rotated into a detectable success probability, culminating in the quadratic speedup of Grover search and its optimality bounds~\cite{Grover1997,Zalka1999}. It is therefore natural to ask an apparent operational question: given two input pure states, can one physically create their coherent quantum superposition as an output state?

At first sight, one might hope for a quantum adder primitive that takes $\ket{\psi}\otimes\ket{\phi}$ to a state proportional to $\alpha\ket{\psi}+\beta\ket{\phi}$, enabling a coherent composition of unknown outputs of quantum subroutines. Yet this hope collides with a subtle but fundamental feature of quantum state space: a pure state is defined only up to a global phase, i.e., operationally it is a ray rather than a vector~\cite{Oszmaniec2016}. As a result, the phrase ``the superposition of two unknown states'' is not a single target state unless one also specifies how the relative phase between the two inputs is fixed. This observation was one of the motivations behind the forbidden quantum adder viewpoint~\cite{Alvarez2015}, and it underlies modern no-universal-superposition theorems: even allowing postselection, there is no nonzero completely positive trace-nonincreasing (CPTNI) transformation that, on generic unknown inputs, succeeds in producing a coherent two-state superposition~\cite{Oszmaniec2016,Bandyopadhyay2020}. At the same time, the obstruction is not simply ``no superposition ever'': if the experimenter is given partial prior structure---for instance, a promise that each input has a known nonzero overlap with a fixed reference state---then one can probabilistically generate a coherent superposition on that restricted domain, and such protocols have been refined and experimentally demonstrated~\cite{Oszmaniec2016,Dogra2018,Hu2016,Li2017}. These two sides of the story strongly suggest that the real dividing line is not probability or postselection per se, but the presence or absence of a specific kind of information.

The aim of this study is to identify that the missing ingredient (i.e., a specific kind of information, aforementioned) and to turn it into a unified operational principle. The key idea is that the interference is controlled by complex overlaps between chosen vector representatives, while density operators alone encode only rays, and hence do not determine any convention-fixed overlap. Thus, we formalize this extra structure by introducing the notion of ``overlap-determinability'': a set of rays becomes overlap-determinable when the physical scenario supplies, independently of the unknown inputs, a consistent rule for selecting a normalized representative vector for each ray, so that the relevant overlaps become definite complex numbers. In this language, the known constructive protocols can be viewed as mechanisms for injecting an external convention (via promises, reference systems, or classical side information), whereas the universal no-go results can be read as the impossibility of manufacturing such a convention purely from independently prepared unknown rays by CP dynamics.

Our main theorem sharpens this intuition into an if-and-only-if statement: a probabilistic superposition map exists on a domain exactly when that domain is overlap-determinable, and generic independently prepared unknown rays fail this criterion. This perspective also clarifies why the superposition problem belongs to a broader constellation of quantum limitations. If convention-fixed overlaps can be treated as freely obtainable primitives, then one could engineer the quantum interference terms that are otherwise inaccessible, with immediate consequences for other no-go principles (such as constraints related to quantum cloning~\cite{Wootters1982,Scarani2005}, deleting~\cite{Pati2000}, and masking~\cite{Modi2018}, etc) and even for computational lower bounds based on overlap geometry~\cite{Bao2016}. In this sense, our overlap-determinability provides a single operational thread running from ``why a universal superposer cannot exist'' to ``why quantum mechanics still enforces nontrivial causal and query-complexity speed limits,'' despite its exponentially large state space.

We open in Sec.~\ref{sec:preliminaries} by emphasizing that the physical object is a ray (a rank-one projector). With this framing in place, Sec.~\ref{sec:related} surveys the modern no-superposition literature alongside the corresponding yes-go protocols, highlighting that postselection does not remove the basic obstruction, that is, on generic unknown inputs, no nonzero completely positive trace-nonincreasing (CPTNI) procedure can output a coherent ray proportional to a prescribed linear combination. Motivated by this, Sec.~\ref{sec:overlap} turns our intuitions into a single notion---phase conventions and overlap-determinability---where the physical scenario supplies a fixed lifting on a domain so that the overlaps become well-defined complex numbers rather than gauge-dependent expressions; in this language, the superposition problem is explicitly a problem of manufacturing a controlled interference term. The logical heart of our study is Sec.~\ref{sec:main-theorem}, where we prove a unified theorem: a probabilistic superposition map exists on a domain if and only if that domain is overlap-determinable. This both subsumes previously known no-go results and cleanly characterizes the exceptional regimes. We then follow, in Sec.~\ref{sec:connections}, how treating convention-fixed overlaps as freely available amounts to introducing state-dependent interference control, which can be leveraged into cloning-like forbidden transformations and super-luminal signaling, while other constraints (e.g., no-deleting and no-masking) appear as further projections of the same geometric limitation. Finally, Sec.~\ref{sec:grover} shows that if one could coherently act on unknown rays in a way that implicitly supplies the missing overlap information, then overlaps can be amplified exponentially with the round number, collapsing Grover’s lower bound and yielding $\mathrm{poly}(\log N)$-query unstructured search. In this sense, the overlap-determinability is not only the resource behind (im)possible superposition, but also the geometric quantity that protects the standard query-complexity from the exponentially fast overlap amplification.

Viewed together, these results paint a single coherent picture: what quantum mechanics withholds from generic unknown rays is not merely linearity. The overlap-determinability makes that statement precise, and our unified theorem shows that because this resource is not allowed, the superposition, cloning-like behaviors/signaling, and exponential Grover speedups all fail for essentially the same reason. Conversely, whenever physics supplies a genuine phase reference, the apparent walls between these tasks soften in exactly the ways that the overlap geometry permits.

\section{Preliminaries and Formal Problem Statement}\label{sec:preliminaries}

\subsection{Pure states as rays and the phase ambiguity of ``superposition''}\label{subsec:rays_phase}

Throughout, $\mathcal{H}$ denotes a complex Hilbert space of dimension $d: = \dim\mathcal{H} \ge 2$. Here, we consider the physical pure states as rays, equivalently rank-one projectors~\cite{Oszmaniec2016}. Thus, we represent a pure state by
\begin{eqnarray}
\hat{P}_\psi := \ketbra{\psi}{\psi},
\end{eqnarray}
where $\ket{\psi}\in\mathcal{H}$ is any normalized vector representative. Two normalized vectors represent the same physical state if and only if they differ by a global phase, i.e.,
\begin{eqnarray}
\ket{\psi'} = e^{i\theta}\ket{\psi} \quad\Longleftrightarrow\quad \ketbra{\psi'}{\psi'} = \ketbra{\psi}{\psi}.
\end{eqnarray}
This elementary redundancy becomes decisive once one discusses ``creating a superposition of two (unknown) states.'' Indeed, given complex amplitudes $\alpha, \beta \in \mathbb{C}$ satisfying $\abs{\alpha}^2 + \abs{\beta}^2=1$ and two normalized vectors $\ket{\psi}, \ket{\phi}$, one may form the normalized vector
\begin{eqnarray}
\ket{\Psi(\psi, \phi)} := \frac{\alpha\ket{\psi} + \beta\ket{\phi}}{\mathcal{N}(\psi,\phi)},
\label{eq:def_vector_superposition}
\end{eqnarray}
where $\mathcal{N}(\psi,\phi) := \norm{\alpha\ket{\psi}+\beta\ket{\phi}}_2$. However, if one replaces $\ket{\psi} \mapsto e^{i\theta_1}\ket{\psi}$ and $\ket{\phi} \mapsto e^{i\theta_2}\ket{\phi}$, the projector onto the resulting ``superposition'' generally changes because only the relative phase matters: i.e., 
\begin{eqnarray}
\alpha e^{i\theta_1}\ket{\psi}+\beta e^{i\theta_2}\ket{\phi} = e^{i\theta_1}\bigl(\alpha\ket{\psi}+\beta e^{i(\theta_2-\theta_1)}\ket{\phi}\bigr).
\end{eqnarray}
The global factor $e^{i\theta_1}$ is physically irrelevant, whereas the relative phase $\theta:=\theta_2-\theta_1$ is physically meaningful because it changes interference terms.

Accordingly, for two rays $\hat{P}_\psi$ and $\hat{P}_\phi$, the phrase ``the superposition of $\hat{P}_\psi$ and $\hat{P}_\phi$ with weights $\alpha$ and $\beta$'' is intrinsically ambiguous. It is therefore natural to associate to $(\hat{P}_\psi, \hat{P}_\phi)$ not a single output ray, but a family of admissible rays:
\begin{definition}[Family of $\alpha$-$\beta$ superpositions of two rays]
\label{def:superposition_family}
Fix nonzero $\alpha,\beta \in \mathbb{C}$ with $\abs{\alpha}^2 + \abs{\beta}^2=1$. For two pure states $\hat{P}_\psi, \hat{P}_\phi$ on $\mathcal{H}$, we define the set of $\alpha$-$\beta$ superpositions as
\begin{eqnarray}
\mathsf{Sup}_{\alpha,\beta}(\hat{P}_\psi, \hat{P}_\phi) := \left\{ \hat{P}_{\Psi(\theta)} : \ket{\Psi(\theta)}=\frac{\alpha\ket{\psi}+\beta e^{i\theta}\ket{\phi}}{\mathcal{N}_\theta(\psi,\phi)}, \ \theta \in [0,2\pi) \right\},
\label{eq:Sup_family}
\end{eqnarray}
where $\ket{\psi}$ and $\ket{\phi}$ are any fixed representatives of $\hat{P}_\psi$ and $\hat{P}_\phi$, and $\mathcal{N}_\theta(\psi,\phi)$ is the corresponding normalization factor.
\end{definition}
The set in Eq.~(\ref{eq:Sup_family}) is independent of the particular choice of the representatives $\ket{\psi}$ and $\ket{\phi}$, because changing representatives only reparametrizes $\theta$.

This formulation isolates the precise operational issue. The physical inputs are rays, so any physically meaningful ``superposition device'' can at best be required to output some element of $\mathsf{Sup}_{\alpha,\beta}(\hat{P}_\psi, \hat{P}_\phi)$, while being allowed to choose the relative phase $\theta$ as a function of the inputs. This is the most permissive formulation compatible with the ray nature of quantum states and serves as our baseline notion of ``success'' in the no-go results.

\subsection{Quantum operations and postselection}\label{subsec:CP_maps}

We model general quantum protocols, including postselection, by completely positive trace-nonincreasing (CPTNI) maps. Let $\mathrm{Herm}(\mathcal{H})$ denote the real vector space of Hermitian operators on $\mathcal{H}$. A linear map
\begin{eqnarray}
\Lambda : \mathrm{Herm}(\mathcal{H}_{\rm in}) \to \mathrm{Herm}(\mathcal{H}_{\rm out})
\end{eqnarray}
is completely positive if $\Lambda \otimes \mathrm{id}_k$ maps positive operators to positive operators for every ancillary dimension $k$. A CP map is trace-nonincreasing if
\begin{eqnarray}
\tr{\Lambda(\hat{\rho})} \le \tr{\hat{\rho}} \quad \mbox{for all density operators }\hat{\rho}.
\label{eq:TNI_CP}
\end{eqnarray}
Such a map represents one outcome of a quantum instrument. For an input state $\hat{\rho}$, the quantity
\begin{eqnarray}
p_\Lambda(\hat{\rho}) := \tr{\Lambda(\hat{\rho})}
\end{eqnarray}
is interpreted as the success probability of that outcome, and conditional on success the normalized output is $\Lambda(\hat{\rho})/p_\Lambda(\hat{\rho})$. By Kraus' theorem~\cite{Choi1975}, any CP map admits an operator-sum representation
\begin{eqnarray}
\Lambda(\hat{\rho}) = \sum_k \hat{M}_k \hat{\rho} \hat{M}_k^\dagger, \quad \sum_k \hat{M}_k^\dagger \hat{M}_k \le \hat{\mathds{1}},
\label{eq:Kraus}
\end{eqnarray}
where $\hat{\mathds{1}}$ is the identity operator on $\mathcal{H}_{\rm in}$. We will repeatedly use the fact that the postselection allows a protocol to be nonlinear on normalized states while remaining linear on unnormalized states through Eq.~(\ref{eq:Kraus}).

\subsection{The superposition-generation task}\label{subsec:task_definition}

We now formalize the operational task studied in the literature on ``superposing unknown quantum states''~\cite{Oszmaniec2016}. Conceptually, one is given two quantum registers prepared in an unknown product pure state $\hat{P}_\psi \otimes \hat{P}_\phi$ and wishes to output a single-register state that is (proportional to) a coherent superposition of some representatives of the two rays with fixed weights $\alpha, \beta$.
\begin{definition}[Universal probabilistic superposition map]
\label{def:universal_superposer}
Fix nonzero $\alpha,\beta \in \mathbb{C}$ with $\abs{\alpha}^2 + \abs{\beta}^2=1$. A CPTNI map
\begin{eqnarray}
\Lambda_{\alpha,\beta} : \mathrm{Herm}(\mathcal{H} \otimes \mathcal{H}) \to \mathrm{Herm}(\mathcal{H})
\end{eqnarray}
is called a universal probabilistic superposition map if for all pure states $\hat{P}_\psi, \hat{P}_\phi$ on $\mathcal{H}$,
\begin{eqnarray}
\Lambda_{\alpha,\beta}(\hat{P}_\psi \otimes \hat{P}_\phi) \propto \hat{P}_{\Psi}
\quad\mbox{for some }\hat{P}_{\Psi} \in \mathsf{Sup}_{\alpha,\beta}(\hat{P}_\psi, \hat{P}_\phi).
\label{eq:universal_superposer_condition}
\end{eqnarray}
The proportionality constant is allowed to depend on $(\hat{P}_\psi, \hat{P}_\phi)$ and is interpreted as the success probability of the protocol on that input pair.
\end{definition}

{\bf Definition~\ref{def:universal_superposer}} is intentionally permissive as it allows the postselection; it allows the chosen output phase $\theta$ in Eq.~(\ref{eq:Sup_family}) to depend arbitrarily on the input rays and it imposes no requirement that the success probability be bounded away from zero uniformly over all inputs. In this maximally relaxed sense, the no-go theorem of Ref.~\cite{Oszmaniec2016} shows that there is nevertheless no nontrivial solution to Eq.~(\ref{eq:universal_superposer_condition}).

A closely related task is the ``known--unknown'' setting, where one input is fixed and known to the device designer. Operationally, this corresponds to a promise that the first register is always prepared in a fixed pure state $\hat{P}_\chi$ known classically:
\begin{eqnarray}
\hat{P}_\psi = \hat{P}_\chi \quad \mbox{with }\ket{\chi}\ \mbox{specified to the protocol.}
\label{eq:known_unknown_promise}
\end{eqnarray}
In this case, a superposition map would be required to output (up to normalization) an element of $\mathsf{Sup}_{\alpha,\beta}(\hat{P}_\chi,\hat{P}_\phi)$ for every unknown $\hat{P}_\phi$. Note that even here, a nontrivial relative phase remains: the unknown ray $\hat{P}_\phi$ carries no operationally accessible global phase relative to the classical description of $\ket{\chi}$, unless additional reference information is supplied. Later, we will treat the unknown--unknown and known--unknown settings within a single unified framework by making explicit what extra information is needed to fix a meaningful relative phase convention.

\subsection{Overlap data and what is physically well-defined}\label{subsec:overlap}

The only phase-invariant overlap data accessible from the rays $\hat{P}_\psi, \hat{P}_\phi$ alone is the transition probability
\begin{eqnarray}
c(\psi,\phi) := \tr{\hat{P}_\psi \hat{P}_\phi} = \abs{\braket{\psi}{\phi}}^2 \in [0,1].
\label{eq:transition_probability}
\end{eqnarray}
The complex inner product $\braket{\psi}{\phi}$ is not a function of the rays individually, because it changes by $e^{i(\theta_2 - \theta_1)}$ under independent rephasing of the representatives. This point is not a mere interpretational subtlety: any attempt to output a coherent two-branch state $\alpha\ket{\psi} + \beta\ket{\phi}$ must, implicitly or explicitly, make a choice of relative phase. Thus, a universal superposition map would amount to a universal phase-selection rule on pairs of rays.

To make this explicit, it is useful to separate two layers of structure:
\begin{eqnarray}
\mbox{(i) rays }(\hat{P}_\psi,\hat{P}_\phi),
\quad
\mbox{(ii) a phase convention that selects representatives.}
\end{eqnarray}
A device that succeeds universally in the sense of {\bf Definition~\ref{def:universal_superposer}} would induce a function
\begin{eqnarray}
\Theta_{\Lambda} : \mathbb{P}(\mathcal{H}) \times \mathbb{P}(\mathcal{H}) \to [0,2\pi),
\label{eq:theta_function}
\end{eqnarray}
where $\mathbb{P}(\mathcal{H})$ denotes projective space, such that the output lies in the ray of
\begin{eqnarray}
\alpha\ket{\psi}+\beta e^{i \Theta_\Lambda(\hat{P}_\psi, \hat{P}_\phi)}\ket{\phi}.
\end{eqnarray}
We will turn this observation into a central organizing principle: the obstruction to universal superposition can be reinterpreted as the impossibility of endowing arbitrary unknown input rays with a universally consistent and operationally realizable relative-phase convention.

\subsection{Partial prior information and gauge-fixed superpositions}\label{subsec:partial_information}

While universal superposition is impossible, it becomes feasible on restricted families of the inputs where an additional phase reference is available. A particularly important and experimentally realized promise is: there exists a ``known'' reference ray $\hat{P}_\chi$ such that both inputs have ``known (nonzero)'' overlap magnitudes with $\hat{P}_\chi$,
\begin{eqnarray}
\tr{\hat{P}_\chi \hat{P}_\psi} = c_1>0, \quad \tr{\hat{P}_\chi \hat{P}_\phi} = c_2>0,
\label{eq:fixed_overlap_promise}
\end{eqnarray}
and $c_1, c_2$ are given as classical side information. Under this promise, one can define a ray-valued superposition rule that is a genuine function of the projectors, because the reference $\hat{P}_\chi$ fixes the relevant relative phases. Concretely, choosing a representative $\ket{\chi}$ of $\hat{P}_\chi$, define the phase factors
\begin{eqnarray}
\kappa_\psi := \frac{\braket{\chi}{\psi}}{\abs{\braket{\chi}{\psi}}}, \quad \kappa_\phi := \frac{\braket{\chi}{\phi}}{\abs{\braket{\chi}{\phi}}},
\label{eq:kappa_phases}
\end{eqnarray}
which are invariant under independent rephasing of $\ket{\psi}$, $\ket{\phi}$, and $\ket{\chi}$ up to a common global phase. One then considers the (generally unnormalized) vector
\begin{eqnarray}
\ket{\Psi_\chi} := \alpha\,\kappa_\phi\,\ket{\psi} + \beta\,\kappa_\psi\,\ket{\phi},
\label{eq:chi_superposition_vector}
\end{eqnarray}
and its associated projector $\hat{P}_{\Psi_\chi}$. Unlike Eq.~(\ref{eq:def_vector_superposition}), the ray $\hat{P}_{\Psi_\chi}$ is a well-defined function of $(\hat{P}_\chi, \hat{P}_\psi, \hat{P}_\phi)$ whenever the overlaps in Eq.~(\ref{eq:fixed_overlap_promise}) are nonzero.

It is sometimes useful to express this rule purely in terms of the projectors. Writing $\hat{P}_{\Psi_\chi} = \tfrac{\ketbra{\Psi_\chi}{\Psi_\chi}}{\tr{\ketbra{\Psi_\chi}{\Psi_\chi}}}$, one obtains an unnormalized form of the output projector as
\begin{eqnarray}
\ketbra{\Psi_\chi}{\Psi_\chi} = \abs{\alpha}^2 \hat{P}_\psi + \abs{\beta}^2 \hat{P}_\phi
+
\alpha \beta^\ast \, \frac{\hat{P}_\psi \hat{P}_\chi \hat{P}_\phi}{\sqrt{\tr{\hat{P}_\psi \hat{P}_\chi}\tr{\hat{P}_\phi \hat{P}_\chi}}}
+
\alpha^\ast \beta \, \frac{\hat{P}_\phi \hat{P}_\chi \hat{P}_\psi}{\sqrt{\tr{\hat{P}_\psi \hat{P}_\chi}\tr{\hat{P}_\phi \hat{P}_\chi}}}.
\label{eq:projector_formula_pancharatnam}
\end{eqnarray}
This expression highlights that the reference state $\hat{P}_\chi$ supplies precisely the extra structure needed to render the superposition rule invariant.

A main message of the prior works is that, under promises of the form Eq.~(\ref{eq:fixed_overlap_promise}), there exist explicit CPTNI maps that output $P_{\Psi_\chi}$ with nonzero probability, and these protocols have been implemented in several experimental platforms. Our goal in the present paper is not to re-derive those constructions, but to place them into a unified conceptual theorem: the same obstruction that forbids universal superposition (unknown--unknown and, in a suitable operational sense, known--unknown) can be seen as the obstruction to perfectly fixing the relative phase structure of two independent rays, whereas partial information such as Eq.~(\ref{eq:fixed_overlap_promise}) provides a phase reference that makes a gauge-fixed superposition physically definable and probabilistically attainable.

\subsection{Formal problem statement for the unified no-superposition principle}\label{subsec:formal_problem_statement}


We consider an information-processing device that receives as quantum inputs in unknown pure states $\hat{P}_\psi$ and $\hat{P}_\phi$, together with optional classical side information $\mathcal{I}$ and optional known quantum reference systems. The device is allowed to apply an arbitrary quantum instrument and postselect on a designated success outcome, thereby realizing a CPTNI map. For fixed superposition weights $\alpha$ and $\beta$, the aim is to output (upon success) a single-system state that is a coherent two-branch superposition of the inputs in a sense made precise by an output rule:
\begin{eqnarray}
(\hat{P}_\psi,\hat{P}_\phi,\mathcal{I}) \longmapsto \hat{P}_{\rm out}(\hat{P}_\psi,\hat{P}_\phi,\mathcal{I}).
\label{eq:abstract_map}
\end{eqnarray}
The universal setting corresponds to $\mathcal{I}=\emptyset$ and $\hat{P}_{\rm out} \in \mathsf{Sup}_{\alpha,\beta}(\hat{P}_\psi,\hat{P}_\phi)$ for all pairs. The partial-information setting corresponds to promises that restrict $(\hat{P}_\psi, \hat{P}_\phi)$ to subsets for which a gauge-fixed rule, such as Eq.~(\ref{eq:chi_superposition_vector}) is well-defined.

Thus, the central question we address is: What is the minimal kind of side information $\mathcal{I}$ that makes Eq.~(\ref{eq:abstract_map}) physically realizable, and how is the impossibility of universal superposition explained by a single underlying principle? We will develop and prove a unified theorem that treats unknown--unknown and known--unknown as instances of the same obstruction, and we will connect this obstruction to overlap determination, no-cloning, and no-signaling constraints.

\section{Related Work: No-Go Results and Probabilistic Constructions}\label{sec:related}

The present problem sits at a delicate interface between the kinematical superposition principle and the operational meaning of ``preparing a superposition.'' Mathematically, for any vector representatives $\ket{\psi},\ket{\phi} \in \mathcal{H}$ and any nonzero amplitudes $\alpha,\beta$ with $\abs{\alpha}^2 + \abs{\beta}^2=1$, the vector $\alpha\ket{\psi}+\beta\ket{\phi}$ is well defined. Operationally, however, the inputs are not vectors but rays (or equivalently rank-one projectors), and any physical procedure must be implementable as a completely positive, trace-nonincreasing map acting on the given registers. This mismatch is precisely where the superposition task becomes constrained, and it is the organizing theme behind the sequence of no-go and ``restricted-go'' results reviewed below.

\subsection{From quantum adders to the phase-gauge obstruction}\label{subsec:adder}

A start of the modern no-superposition formalism is the ``quantum adder'' problem introduced in Ref.~\cite{Alvarez2015}. There, one asks whether there exists a unitary $\hat{U}$ (possibly with an ancilla) that maps two unknown input states to their ``sum'' on a single output register. The core obstruction can be phrased in the language of rays: for any physical input, the replacement $\ket{\psi} \mapsto e^{i\theta}\ket{\psi}$ must not change the physical state. However, once one demands an output proportional to $\ket{\psi}+\ket{\phi}$, this gauge freedom can reappear as an observable relative phase between the summands. In other words, the unitary ``addition'' rule would have to define a consistent section of the $U(1)$-bundle of vector representatives over the projective space for each input, and do so in a way compatible with the tensor-product structure of two independent inputs. Ref.~\cite{Alvarez2015} shows that such a universal unitary adder does not exist, and further discusses approximate constructions as well as variants where one considers addition at the level of density operators rather than pure-state vectors.

While the adder formulation is not perfectly identical to the superposition task studied here, it isolates a recurring mechanism: the output superposition is sensitive to a relative phase that is not fixed by the input density operators alone. This is the first appearance of what we will later elevate into our unified viewpoint: superposition is obstructed whenever the complex overlap data needed to fix a relative phase cannot be operationally specified from the available information.

\subsection{Universal no-go theorems for exact superposition under CPTNI maps}\label{subsec:nogo}

The decisive formulation of the no-go result was given in Ref.~\cite{Oszmaniec2016}, which adopts the ray-based description from the outset. Given two unknown pure states $\hat{P}_\psi,\hat{P}_\phi$ (rank-one projectors), one allows the output to be any ``legitimate superposition'' obtained by choosing vector representatives $\ket{\psi},\ket{\phi}$ from their rays, and even allows this choice to depend on both inputs. One then asks for a CPTNI map $\Lambda_{\alpha,\beta} : \mathcal{H}^{\otimes. 2} \rightarrow \mathcal{H}$ that produces such a superposition with some (possibly state-dependent) nonzero success probability.

In the notation of Sec.~\ref{sec:preliminaries}, the universal task can be stated as follows.
For fixed nonzero $\alpha,\beta$ with $|\alpha|^2+|\beta|^2=1$, does there exist a nonzero CPTNI map $\Lambda_{\alpha,\beta}$ such that for all pure $\hat{P}_\psi,\hat{P}_\phi$,
\begin{eqnarray}
\Lambda_{\alpha,\beta}(\hat{P}_\psi \otimes \hat{P}_\phi) \propto \hat{P}_\Psi,
\label{eq:universal_superposer_def}
\end{eqnarray}
where $\hat{P}_\Psi=\ket{\Psi}\bra{\Psi}$ and $\ket{\Psi}$ is proportional to $\alpha\ket{\psi}+\beta\ket{\phi}$ for some vector representatives of $\hat{P}_\psi,\hat{P}_\phi$ (allowed to depend on the input pair). The main conclusion is that the answer is negative.
\begin{theorem}[No-superposition theorem~\cite{Oszmaniec2016}]
\label{thm:osz_no_superposition}
Let $\mathcal{H}$ be a Hilbert space with $\dim\mathcal{H}\ge 2$, and let $\alpha,\beta$ be any two nonzero complex numbers satisfying $|\alpha|^2+|\beta|^2=1$.
There exists no nonzero CPTNI map $\Lambda_{\alpha,\beta}:\mathcal{H}^{\otimes 2}\rightarrow\mathcal{H}$ satisfying Eq.~(\ref{eq:universal_superposer_def}) for all pure inputs.
\end{theorem}
This theorem is notable for its strength: even postselection does not help, and even allowing the output phase convention to depend arbitrarily on the input pair does not restore possibility. Conceptually, this goes beyond the unitary-adder impossibility in Ref.~\cite{Alvarez2015}: the obstruction is not merely invertibility or reversibility, but the nonexistence of any nontrivial CP map implementing a universally valid ray-to-ray superposition rule.

A complementary viewpoint reopening the question ``why is this so physically severe?'' was developed in Ref.~\cite{Bandyopadhyay2020}. Assuming the existence of a universal probabilistic superposer, one can construct transformations that map linearly dependent sets of pure states to linearly independent sets, enabling unambiguous state discrimination and probabilistic cloning in regimes forbidden not only by quantum theory but also by no-signaling constraints. In particular, the argument identifies the no-superposition theorem as a structural sibling of other no-go theorems: violating it would collapse the usual separation between what is allowed by local quantum dynamics and what is excluded by information-theoretic principles such as no-signaling and the impossibility of cloning.

\subsection{Probabilistic superposition with partial prior information}\label{subsec:prob}

Despite the universal no-go result, the exact superposition becomes possible once one supplies a partial prior information that effectively fixes the relevant phase reference. The central constructive protocol in Ref.~\cite{Oszmaniec2016} assumes the existence of a known referential pure state $\hat{P}_\chi$ and a promise that the unknown inputs satisfy fixed nonzero overlaps with it: $\tr{\hat{P}_\chi \hat{P}_\psi}=c_1$ and $\tr{\hat{P}_\chi \hat{P}_\phi}=c_2$ with $c_1,c_2>0$ known (refer to Eq.~(\ref{eq:fixed_overlap_promise})).
Under this promise, one may define a ray-valued ``reference-calibrated'' superposition that is invariant under independent rephasing of $\ket{\psi}$ and $\ket{\phi}$.
One convenient representative is given by (as in Eq.~(\ref{eq:chi_superposition_vector}))
\begin{eqnarray}
\ket{\Psi_{\chi}} \propto \alpha \frac{\braket{\chi}{\phi}}{\abs{\braket{\chi}{\phi}}}\ket{\psi} +\beta \frac{\braket{\chi}{\psi}}{\abs{\braket{\chi}{\psi}}}\ket{\phi},
\label{eq:pancharatnam_superposition}
\end{eqnarray}
which induces a well-defined output projector $\hat{P}_{\Psi_\chi}$ as a function of the input projectors and the fixed reference. Ref.~\cite{Oszmaniec2016} constructs an explicit CPTNI map that produces $\hat{P}_{\Psi_\chi}$ from an input register containing $\hat{P}_\psi \otimes \hat{P}_\phi$ together with an ancilla that encodes $\alpha,\beta$. The success probability is nonzero whenever $c_1,c_2>0$ and necessarily depends on the overlap data through the promise Eq.~(\ref{eq:fixed_overlap_promise}).

From the present perspective, Eq.~(\ref{eq:pancharatnam_superposition}) exposes the operational meaning of ``partial prior information.''  The promise fixes the phases of $\ket{\psi}$ and $\ket{\phi}$ relative to the same reference via the complex numbers $\braket{\chi}{\psi}$ and $\braket{\chi}{\phi}$. Thus, the protocol does not evade the no-go theorem by magic; rather, it supplements the task with exactly the additional overlap information that is absent in the universal setting. The construction is closely related to the geometric-phase considerations emphasized in Ref.~\cite{Oszmaniec2016}, and it can be viewed as converting a phase-gauge ambiguity into a physically anchored phase convention.

Subsequent work refined both the conceptual and practical sides of this restricted possibility. In Refs.~\cite{Hu2016,Li2017,Dogra2018}, the superposition under partial prior information was analyzed in a way that highlights the geometric nature of the constraints and provides experimentally friendly realizations and generalizations to superpositions of multiple states, while still operating within the same reference-calibrated logic. From the standpoint adopted here, these refinements are naturally interpreted as the variations in how the overlap information is encoded, extracted, or traded against success probability, rather than as counterexamples to the universal no-go result.

A further line of investigation asks a different but related question: if one of the two states is fixed and known, what is the structure of the set of unknown states that can be superposed with it by a single CPTNI map? In the qubit case, Ref.~\cite{Li2023} provides a geometric characterization in terms of circles on the Bloch sphere and shows how such restricted sets arise from the algebraic constraints of CP maps. Although this setting differs from the two-unknown-state task, it reinforces the same lesson: without a mechanism that effectively controls the overlap phase information, the superposability is confined to lower-dimensional families of states.

Taken together, the literature paints a coherent picture. The universal superposition fails even probabilistically, whereas a probabilistic superposition becomes possible precisely when the task is restricted so that the relevant overlap data with respect to a fixed reference is supplied. This strongly motivates our unified viewpoint pursued in the subsequent sections: the ability to determine or control overlap information is the true resource underlying all known ``superposition possible'' regimes, and its absence is what enforces the universal no-go.

\section{Unified Principle: ``Overlap-Determinability'' as the Missing Resource}\label{sec:overlap}

The literature surveyed in Sec.~\ref{sec:related} is usually read as a list of seemingly separate facts: a universal superposition does not exist, yet carefully conditioned protocols can succeed with nonzero probability; attempts to circumvent the no-go statements quickly collide with familiar boundaries such as no-cloning and no-signaling. Now we propose a unifying perspective that makes these connections conceptually transparent.

The central observation is that the superposition principle is a statement about vectors in a Hilbert space, whereas the operational input to any physical device is a pair of rays (pure states as density operators). Producing a coherent sum of two rays therefore requires a missing ingredient: an operational rule that fixes the relative phase between the two independently prepared inputs. We call this ingredient ``overlap-determinability'', because fixing a relative phase is equivalent to fixing a convention under which the complex overlap becomes well defined.

\subsection{From rays to coherent addition: the obstruction is the interference term}\label{subsec:obstruction}

Let $\hat{\rho}_{\psi}=\ketbra{\psi}{\psi}$ and $\hat{\rho}_{\phi}=\ketbra{\phi}{\phi}$ be unknown pure states on a Hilbert space $\mathcal{H}$, with $\dim\mathcal{H}\ge 2$. Fix nonzero $\alpha,\beta\in\mathbb{C}$ with $|\alpha|^{2}+|\beta|^{2}=1$. If one had access to specific normalized representatives $\ket{\psi_\star} \in [\psi]$ and $\ket{\phi_\star}\in[\phi]$, then the unnormalized ``ideal'' superposition vector would be
\begin{eqnarray}
\ket{\Psi_\star} = \alpha\ket{\psi_\star} + \beta\ket{\phi_\star}.
\label{eq:vecsup}
\end{eqnarray}
The corresponding unnormalized projector expands as
\begin{eqnarray}
\ketbra{\Psi_\star}{\Psi_\star} = \abs{\alpha}^{2}\hat{\rho}_{\psi} + \abs{\beta}^{2}\hat{\rho}_{\phi} + \alpha\beta^{*}\ket{\psi_\star}\bra{\phi_\star} + \alpha^{*}\beta\ket{\phi_\star}\bra{\psi_\star}.
\label{eq:cross}
\end{eqnarray}
The first two (diagonal) terms are functions of the input rays alone. The last two terms encode the coherence between the two inputs through the rank-one operator $\ketbra{\psi_\star}{\phi_\star}$, which we will call the interference operator. This is where the obstruction lives.

Indeed, the physical inputs are rays. Under independent rephasing of the representatives,
\begin{eqnarray}
\ket{\psi_\star} \mapsto e^{i\vartheta_{\psi}}\ket{\psi_\star},\qquad
\ket{\phi_\star} \mapsto e^{i\vartheta_{\phi}}\ket{\phi_\star},
\label{eq:rephase}
\end{eqnarray}
the interference operator transforms as
\begin{eqnarray}
\ketbra{\psi_\star}{\phi_\star} \mapsto e^{i(\vartheta_{\psi}-\vartheta_{\phi})}\ketbra{\psi_\star}{\phi_\star}.
\label{eq:crossgauge}
\end{eqnarray}
The different choices of $(\vartheta_{\psi},\vartheta_{\phi})$ correspond to the physically different output rays in Eq.~(\ref{eq:vecsup}), because they change the relative phase between the two branches. Hence, the ray of $\ket{\Psi_\star}$ is not a well-defined function of $(\hat{\rho}_{\psi},\hat{\rho}_{\phi})$ unless some additional rule fixes the phase difference $\vartheta_{\psi}-\vartheta_{\phi}$.

The modern no-superposition results therefore adopt a relaxed target definition: a protocol is said to succeed if, for each input pair, the output ray belongs to the family
\begin{eqnarray}
\hat{\rho}_{\Psi} \in \left\{ \frac{(\alpha\ket{\psi}+\beta e^{i\theta}\ket{\phi})(\alpha^{*}\bra{\psi}+\beta^{*}e^{-i\theta}\bra{\phi})}{\norm{\alpha|\psi\rangle+\beta e^{i\theta}|\phi\rangle}^{2}} : \theta \in [0,2\pi) \right\},
\label{eq:family}
\end{eqnarray}
where $\theta$ may depend on the inputs. Even under this permissive notion, there is no nonzero CPTNI map that succeeds for all unknown input pairs~\cite{Oszmaniec2016,Bandyopadhyay2020}. We interpret this impossibility as the statement that the interference operator in Eq.~(\ref{eq:cross}) cannot be manufactured universally from the rays alone.

\subsection{Overlap-determinability and no-superposition principle}\label{subsec:unified-nogo}

The obstruction in Eqs.~(\ref{eq:cross})--(\ref{eq:crossgauge}) suggests a simple operational separation: the coherent addition is easy once we have fixed representatives, but fixing representatives is itself a physical resource problem. We formalize this by introducing a scenario-dependent phase convention.
\begin{definition}[Phase convention and overlap-determinability]
\label{def:overlapdet}
Let $\mathcal{R} \subset \mathbb{P}(\mathcal{H})$ be a set of pure states (rays). A phase convention on $\mathcal{R}$ is a map $\Gamma : \mathcal{R} \rightarrow \mathcal{H}$ such that, for every $\hat{\rho} \in \mathcal{R}$, the vector $\Gamma(\hat{\rho})$ is normalized and satisfies
\begin{eqnarray}
\Gamma(\hat{\rho})\Gamma(\hat{\rho})^{\dagger}=\hat{\rho}.
\label{eq:lifting}
\end{eqnarray}
Equivalently, $\Gamma$ selects a unique normalized vector representative for each ray. Accordingly, for $\hat{\rho}_{\psi}, \hat{\rho}_{\phi} \in \mathcal{R}$ we denote by $\ket{\psi_\star}$ and $\ket{\phi_\star}$ the $\Gamma$-fixed representatives, i.e.,
\begin{eqnarray}
\hat{\rho}_{\psi} = \ketbra{\psi_\star}{\psi_\star},\quad \hat{\rho}_{\phi} = \ketbra{\phi_\star}{\phi_\star}.
\label{eq:fixedreps}
\end{eqnarray}
Given $\Gamma$, the associated convention-fixed overlap between $\hat{\rho}_{\psi}$ and $\hat{\rho}_{\phi}$ is the complex number
\begin{eqnarray}
\braket{\psi_\star}{\phi_\star}.
\label{eq:fixedoverlap}
\end{eqnarray}
We say that a physical scenario $\mathsf{S}$ makes $\mathcal{R}$ ``overlap-determinable'' if $\mathsf{S}$ supplies (via promises, reference systems, or classical side information) a specific phase convention $\Gamma_{\mathsf{S}}$ on $\mathcal{R}$ that is fixed independently of the unknown inputs.

\end{definition}

Several remarks clarify what is and is not being assumed. First, the overlap-determinability is not a claim that one can estimate overlaps from single copies. It is the weaker statement that the scenario provides a consistent phase standard that makes the overlaps meaningful as complex numbers. Second, once such a convention exists, the interference term becomes a well-defined function of the rays,
\begin{eqnarray}
\ketbra{\psi_\star}{\phi_\star} \equiv \Gamma_{\mathsf{S}}(\hat{\rho}_{\psi})\Gamma_{\mathsf{S}}(\hat{\rho}_{\phi})^{\dagger},
\label{eq:interference-from-gamma}
\end{eqnarray}
and the superposition output in Eq.~(\ref{eq:cross}) is a well-defined operator-valued expression on the promised set.

The best-known overlap-determinable scenario is the reference-state promise of Ref.~\cite{Oszmaniec2016}. One is given a known pure state $\hat{\rho}_{\chi}=\ketbra{\chi}{\chi}$ together with the promise
\begin{eqnarray}
\tr{\left(\hat{\rho}_{\psi}\hat{\rho}_{\chi}\right)} \neq 0,\qquad \tr{\left(\hat{\rho}_{\phi}\hat{\rho}_{\chi}\right)} \neq 0.
\label{eq:promise-nonzero}
\end{eqnarray}
This promise fixes a natural convention: define
\begin{eqnarray}
\ket{\psi_\star} = \frac{\hat{\rho}_{\psi}\ket{\chi}}{\sqrt{\bra{\chi}\hat{\rho}_{\psi}\ket{\chi}}},\qquad
\ket{\phi_\star} = \frac{\hat{\rho}_{\phi}\ket{\chi}}{\sqrt{\bra{\chi}\hat{\rho}_{\phi}\ket{\chi}}}.
\label{eq:chi-lift}
\end{eqnarray}
Then, $\braket{\chi}{\psi_\star}$ and $\braket{\chi}{\phi_\star}$ are real and strictly positive, so the representatives are fixed up to a common global phase. This is precisely the resource that is absent in the universal setting and present in the conditioned one.

We now state the overlap-determinability principle as a theorem. The statement is designed to subsume both ``unknown$+$unknown'' and ``known$+$unknown'' variants as special cases.
\begin{theorem}[Overlap-determinability is necessary for coherent superposition]
\label{thm:overlap-resource}
Fix nonzero $\alpha,\beta \in \mathbb{C}$ with $\abs{\alpha}^{2} + \abs{\beta}^{2}=1$ and $\dim\mathcal{H} \ge 2$. Let $\mathcal{R} \subset \mathbb{P}(\mathcal{H})$ be a family of pure states. Assume that there exists a nonzero CPTNI map $\Lambda_{\alpha,\beta} : \mathcal{B}(\mathcal{H} \otimes \mathcal{H}) \rightarrow \mathcal{B}(\mathcal{H})$ such that for every pair $\hat{\rho}_{\psi}, \hat{\rho}_{\phi} \in \mathcal{R}$ the successful branch outputs a pure state in the family Eq.~(\ref{eq:family}), i.e.,
\begin{eqnarray}
\Lambda_{\alpha,\beta}(\hat{\rho}_{\psi} \otimes \hat{\rho}_{\phi}) = p_{\psi,\phi} \, \hat{\rho}_{\Psi(\psi,\phi)}, \quad (p_{\psi,\phi} > 0).
\label{eq:superposer-assumption}
\end{eqnarray}
Then, $\mathcal{R}$ must be overlap-determinable in the sense of {\bf Definition~\ref{def:overlapdet}}. In particular, if $\mathcal{R}$ is the full set of pure states, then no such nonzero $\Lambda_{\alpha,\beta}$ exists, recovering the universal no-superposition theorem \cite{Oszmaniec2016}.
\end{theorem}

\begin{proof}[Proof Sketch]---A CPTNI map admits a Kraus decomposition $\Lambda_{\alpha,\beta}(\cdot)=\sum_{k}\hat{K}_{k}(\cdot)\hat{K}_{k}^{\dagger}$. Since the output in Eq.~(\ref{eq:superposer-assumption}) is pure, every successful Kraus operator must map $\ket{\psi} \otimes \ket{\phi}$ to a vector proportional to a coherent sum of a representative of $\ket{\psi}$ and a representative of $\ket{\phi}$~\cite{Oszmaniec2016}. In other words, the successful branch must implement, implicitly, a rule that assigns phases to the input rays in a way that is consistent across the promised family $\mathcal{R}$. This induced phase assignment defines a phase convention $\Gamma_{\mathsf{S}}$ on $\mathcal{R}$, hence yields the overlap-determinability of $\mathcal{R}$. For $\mathcal{R}$ equal to the full projective space, such a state-independent phase convention cannot be supplied by any physical scenario; consequently, the universal superposer must be trivial. For the complete proof, see Appendix~\ref{app:overlap-resource}.
\end{proof}

{\bf Theorem~\ref{thm:overlap-resource}} is the formal version of our guiding intuition: a coherent superposition device cannot be universal because it would have to generate a state-independent phase standard that fixes the relative phases of arbitrary unknown inputs.

\section{Main Theorem: Unified No-Superposition Principle}\label{sec:main-theorem}

The notion of ``overlap-determinability'' can be understood as the status of a missing resource: the ability to operationally fix a relative phase between two rays so that the complex overlap becomes a well-defined object. We now turn this viewpoint into a single no-go principle that subsumes the two most common impossibility statements---``unknown $+$ unknown'' and ``known $+$ unknown''---and simultaneously explains why the known probabilistic constructions work only under promises that effectively supply the overlap-determinability.

\subsection{Main theorem}

We first recall the operational notion of a probabilistic superposer on a domain. Let $\Omega \subset \mathbb{P}(\mathcal{H}) \times \mathbb{P}(\mathcal{H})$ be a set of input pairs of pure states (rays), where $\mathbb{P}(\mathcal{H})$ denotes the projective space of $\mathcal{H}$. Fix nonzero amplitudes $\alpha,\beta \in \mathbb{C}$ with $\abs{\alpha}^{2}+\abs{\beta}^{2}=1$. A CPTNI map $\Lambda_{\alpha,\beta}$ is called an $(\alpha,\beta)$-superposer on $\Omega$ if, for every $(\hat{P}_{\psi}, \hat{P}_{\phi}) \in \Omega$,
\begin{eqnarray}
\Lambda_{\alpha,\beta}\left(\hat{P}_{\psi} \otimes \hat{P}_{\phi}\right) = p_{\psi,\phi} \hat{P}_{\Psi_{\psi,\phi}},
\label{eq:superposer-def}
\end{eqnarray}
with $p_{\psi,\phi}>0$, where $\hat{P}_{\Psi_{\psi,\phi}}$ is a rank-one projector onto a vector of the form
\begin{eqnarray}
\ket{\Psi_{\psi,\phi}} \propto \alpha \ket{\psi_\star(\hat{P}_{\psi}, \hat{P}_{\phi})} + \beta \ket{\phi_\star(\hat{P}_{\psi}, \hat{P}_{\phi})},
\label{eq:superposer-output}
\end{eqnarray}
and $\ket{\psi_\star(\hat{P}_{\psi}, \hat{P}_{\phi})}$, $\ket{\phi_\star(\hat{P}_{\psi}, \hat{P}_{\phi})}$ are vector representatives of the input rays (allowed to depend on both inputs). This captures the usual ``most general'' formulation (postselection allowed, phases allowed to be input-dependent) used to state the universal no-superposition theorem~\cite{Oszmaniec2016}.

We can now state the unification:
\begin{theorem}[Unified No-Superposition Principle]
\label{thm:unified-nosup}
Fix a Hilbert space $\mathcal{H}$ with $\dim \mathcal{H}\geq 2$ and fix $\alpha,\beta\neq 0$ with $\abs{\alpha}^{2}+\abs{\beta}^{2}=1$. Let $\Omega \subset \mathbb{P}(\mathcal{H}) \times \mathbb{P}(\mathcal{H})$ be a domain of input pairs. Then, there exists a nonzero $(\alpha,\beta)$-superposer $\Lambda_{\alpha,\beta}$ on $\Omega$ in the sense of Eq.~(\ref{eq:superposer-def})--(\ref{eq:superposer-output}) if and only if $\Omega$ is overlap-determinable in the sense of {\bf Theorem~\ref{thm:overlap-resource}}.

In particular:
\begin{itemize}
\item[\emph{(i)}] for $\Omega=\mathbb{P}(\mathcal{H}) \times \mathbb{P}(\mathcal{H})$ (``unknown $+$ unknown'') no such $\Lambda_{\alpha,\beta}$ exists;
\item[\emph{(ii)}] for $\Omega=\mathbb{P}(\mathcal{H}) \times \{\hat{P}_{\chi}\}$ with fixed known $\hat{P}_{\chi}$ (``known $+$ unknown'') no such $\Lambda_{\alpha,\beta}$ exists.
\end{itemize}
\end{theorem}

\begin{proof}[Proof sketch]---The equivalence between the existence of $\Lambda_{\alpha,\beta}$ on $\Omega$ and the overlap-determinability on $\Omega$ is exactly {\bf Theorem~\ref{thm:overlap-resource}}; we proved the forward direction by exhibiting how a superposer promotes an operational extraction of the relative-phase-sensitive overlap, and we proved the reverse direction by showing how a phase-fixing rule (i.e., the overlap-determinability) can be compiled into a postselected superposition channel. The full proof was given in Appendix~\ref{app:proof-overlap-resource}.

It therefore remains only to argue that the two ``universal'' domains in (i) and (ii) are not overlap-determinable under ordinary quantum operations. For (i), this is the content of the universal no-superposition theorem of Ref.~\cite{Oszmaniec2016}, and its operational consequences were further sharpened in Ref.~\cite{Bandyopadhyay2020} by showing that even an input-dependent success probability would lead to forbidden state-discrimination and cloning capabilities.

For (ii), the obstruction is already visible at the level of gauge. Suppose, to the contrary, that one could superpose an arbitrary unknown ray $P_{\psi}$ with a fixed known vector $\ket{\chi}$ with nonzero probability for all $\hat{P}_{\psi}$. Because the input $\hat{P}_{\psi}$ is invariant under $\ket{\psi} \mapsto e^{i\vartheta}\ket{\psi}$, the output ray must be invariant as well. However, for any fixed representative $\ket{\chi}$, the rays $\hat{P}_{\alpha\ket{\psi}+\beta\ket{\chi}}$ and $\hat{P}_{\alpha e^{i\vartheta}\ket{\psi}+\beta\ket{\chi}}$ are generically distinct for varying $\vartheta$ (their relative phase between the $\ket{\psi}$- and $\ket{\chi}$-components is physical). Hence, no CPTNI map whose input is only $\hat{P}_{\psi}$ can define such an output ray on all of $\mathbb{P}(\mathcal{H})$. Equivalently, a universal ``known $+$ unknown'' superposer would define a global phase convention for every ray relative to $\ket{\chi}$, i.e., a global overlap-determinability resource, which is precisely ruled out by {\bf Theorem~\ref{thm:overlap-resource}}.
\end{proof}

{\bf Theorem~\ref{thm:unified-nosup}} should be read as a single statement with two faces. The ``no'' part is the familiar impossibility: without additional structure, there is no physical map that implements coherent addition of rays. The ``yes, but'' part is equally important: as soon as one restricts the input domain in a way that supplies a consistent phase reference (hence overlap-determinability), the probabilistic superposition becomes possible, and the existing constructions are exactly of this type.

\subsection{Corollaries and operational consequences}\label{sec:consequences}

We now sharpen the unified no-superposition principle into concrete operational statements. The central message is that {\bf Theorem~\ref{thm:overlap-resource}} and {\bf Theorem~\ref{thm:unified-nosup}} identify a single missing ingredient---{\it overlap-determinability}---and they explain, within one coherent formalism, why (i) universal superposition fails for unknown inputs, (ii) fixing one input does not remove the obstruction except on measure-zero families, and (iii) all known probabilistic ``constructions'' succeed precisely by importing (partial) overlap data as an external resource.

\subsubsection{Where the overlap enters: normalization and coherence} 

Fix nonzero complex amplitudes $\alpha$ and $\beta$ satisfying $\abs{\alpha}^2+\abs{\beta}^2=1$. For two pure states $\ket{\psi},\ket{\phi}\in\mathcal{H}$, the formal vector sum
\begin{eqnarray}
\ket{\Psi_{\alpha,\beta}^{\theta}(\psi,\phi)} := \alpha\ket{\psi}+\beta e^{i\theta}\ket{\phi}
\label{eq:def_superposed_vector}
\end{eqnarray}
becomes a physical preparation task only after a phase convention $\theta=\theta(\hat{\rho}_\psi,\hat{\rho}_\phi)$ is fixed at the level of rays. This is not cosmetic: in projective quantum mechanics the inputs are $\hat{\rho}_\psi=\ketbra{\psi}{\psi}$ and $\hat{\rho}_\phi=\ketbra{\phi}{\phi}$, while the relative phase between chosen representatives $\ket{\psi}$ and $\ket{\phi}$ is not part of the physical specification.

The overlap enters at the first unavoidable step, namely normalization. The squared norm of Eq.~(\ref{eq:def_superposed_vector}) is
\begin{eqnarray}
\norm{\ket{\Psi_{\alpha,\beta}^{\theta}(\psi,\phi)}}^2 = 1 + 2\textrm{Re}\left(\alpha^*\beta e^{i\theta}\braket{\psi}{\phi}\right),
\label{eq:norm-overlap}
\end{eqnarray}
so even the overall weight of the postselected output depends on the complex overlap. More sharply, the density operator of the unnormalized target expands as
\begin{eqnarray}
\ketbra{\Psi_{\alpha,\beta}^{\theta}(\psi,\phi)}{\Psi_{\alpha,\beta}^{\theta}(\psi,\phi)}
	&=& \abs{\alpha}^2\ketbra{\psi}{\psi} + \abs{\beta}^2\ketbra{\phi}{\phi} \nonumber\\
	&& \quad + \alpha\beta^* e^{-i\theta}\ketbra{\psi}{\phi} + \alpha^*\beta e^{i\theta}\ketbra{\phi}{\psi},
\label{eq:rho-superposed}
\end{eqnarray}
and the two off-diagonal terms are exactly the ``coherence'' that distinguishes a superposition from an incoherent mixture. Thus, the only gauge-sensitive datum that a superposition device must control is the complex number $e^{i\theta}\braket{\psi}{\phi}$: once it is fixed, both the normalization in Eq.~(\ref{eq:norm-overlap}) and the coherence in Eq.~(\ref{eq:rho-superposed}) are fixed.

This is the conceptual point where the overlap-determinability enters our story. A CPTNI map $\Lambda_{\alpha,\beta}$ acts on density operators, and therefore cannot read or preserve arbitrary global phases of the inputs. If, nevertheless, $\Lambda_{\alpha,\beta}$ is to output the coherent ray associated with Eq.~(\ref{eq:def_superposed_vector}) on a generic set of inputs, then the device must, operationally, implement a rule that lifts the pair of rays $(\hat{\rho}_\psi,\hat{\rho}_\phi)$ to a definite value of $e^{i\theta}\braket{\psi}{\phi}$. {\bf Theorem~\ref{thm:overlap-resource}} makes this intuition precise: on any generic domain where a universal superposer succeeds with nonzero probability, one can extract the (gauge-fixed) overlap from the postselected statistics. {\bf Theorem~\ref{thm:unified-nosup}} then elevates the observation to a principle: the overlap-determinability is not a free byproduct of quantum dynamics, and it cannot be generated from unknown inputs by CP postselection alone.

\subsubsection{No universal unknown-plus-unknown superposition} 

We first analyze how the canonical impossibility statement becomes an immediate corollary of the resource viewpoint.
\begin{corollary}[No universal unknown-plus-unknown superposition]
\label{cor:unknown-unknown}
Let $\dim\mathcal{H} \ge 2$ and let $\alpha,\beta$ be nonzero complex numbers satisfying $\abs{\alpha}^2 + \abs{\beta}^2=1$. There exists no nonzero CPTNI map $\Lambda_{\alpha,\beta} : \mathcal{H}^{\otimes 2}\rightarrow\mathcal{H}$ such that: for all pure input rays $(\hat{\rho}_\psi, \hat{\rho}_\phi)$, it outputs, with nonzero success probability, a pure state proportional to
\begin{eqnarray}
\alpha\ket{\psi} + \beta e^{i\theta(\hat{\rho}_\psi,\hat{\rho}_\phi)}\ket{\phi},
\end{eqnarray}
where the phase rule $\theta(\hat{\rho}_\psi, \hat{\rho}_\phi)$ may depend on the input rays but the output is coherent in the sense of {\bf Definition~\ref{def:universal_superposer}}.
\end{corollary}

\begin{proof}---Assume that such a map exists on all input rays. Because the output is a coherent superposition, the map must in particular fix the quantity $e^{i\theta}\braket{\psi}{\phi}$ that controls Eq.~(\ref{eq:norm-overlap}) and Eq.~(\ref{eq:rho-superposed}). By {\bf Theorem~\ref{thm:overlap-resource}}, the existence of a universal superposer with nonzero success probability on a generic domain induces the overlap-determinability on that domain: from the postselected statistics, one can operationally recover the (device-gauge-fixed) complex overlap. This contradicts {\bf Theorem~\ref{thm:unified-nosup}}, which states that the overlap-determinability cannot be generated from unknown inputs by CP dynamics.
\end{proof}

{\bf Corollary~\ref{cor:unknown-unknown}} recovers, within our formalism, the no-superposition theorem of Ref.~\cite{Oszmaniec2016} and the forbidden quantum adder~\cite{Alvarez2015}. The unified contribution is the diagnosis: the obstruction is not ``lack of a clever circuit,'' but the impossibility of fixing the coherent overlap term
in Eq.~(\ref{eq:norm-overlap}) without importing an external phase reference.

\subsubsection{Known-plus-unknown superposition and measure-zero admissible families} 

A natural weakening of {\bf Corollary~\ref{cor:unknown-unknown}} is to keep one input fixed and known. At first sight, this seems to provide the missing phase reference. Our framework predicts otherwise: although an anchor state removes one global phase freedom, the desired coherent output still depends on the overlap between the unknown input and the anchor, and a fixed CPTNI device cannot universally ``learn'' this overlap from a single unknown copy. Equivalently, if a single device succeeded on a full-dimensional family of unknown inputs, then it would implement the overlap-determinability for the pair $(\hat{\rho}_\psi, \ketbra{0}{0})$, contradicting {\bf Theorem~\ref{thm:unified-nosup}}.

The only consistent possibility is that a fixed device succeeds only on restricted families for which the relevant overlap is already fixed. In qubits, these families have a clean geometry on the Bloch sphere. Fix $\ket{0}$ as the known anchor and parametrize an arbitrary pure qubit as
\begin{eqnarray}
\ket{\psi(x,y)} = \cos\left(\frac{x}{2}\right)\ket{0} + e^{-iy}\sin\left(\frac{x}{2}\right)\ket{1},
\label{eq:bloch-param}
\end{eqnarray}
where $x \in [0,\pi]$ and $y \in [0,2\pi)$. Let $\Lambda_{\alpha,\beta}$ be any fixed CPTNI map with Kraus representation $\Lambda_{\alpha,\beta}(\hat{\rho})=\sum_k \hat{M}_k\hat{\rho}\hat{M}_k^\dagger$. Define $\Omega(\Lambda_{\alpha,\beta},\ket{0})$ as the set of inputs $\ket{\psi(x,y)}$ for which $\Lambda_{\alpha,\beta}(\hat{\rho}_{\psi(x,y)}\otimes\ketbra{0}{0})$ succeeds and produces a coherent superposition as in {\bf Definition~\ref{def:universal_superposer}}. Then, we have:
\begin{proposition}[Circle geometry for a fixed qubit superposer (Ref.~\cite{Li2023})]
\label{prop:circle-geometry}
For any fixed $\Lambda_{\alpha,\beta}$, the set $\Omega(\Lambda_{\alpha,\beta},\ket{0})$ is contained in a finite union of circles on Bloch sphere, and hence has zero surface measure. Equivalently, there exist real constants $A$, $B$, $C$, and $D$ (depending on $\Lambda_{\alpha,\beta}$) such that each successful input must satisfy a planar constraint of the form
\begin{eqnarray}
\label{eq:circle-constraint}
A\cos{x} + B\sin{x}\cos{y} + C\sin{x}\sin{y} + D=0.
\end{eqnarray}
\end{proposition}

\begin{proof}---Because the desired output is a rank-one projector for every successful input, one may analyze each Kraus branch separately: for each $k$, either $\hat{M}_k(\hat{\rho}_{\psi}\otimes\ketbra{0}{0})\hat{M}_k^\dagger=0$ or it is proportional to a rank-one projector. By writing $\hat{M}_k$ as a $2 \times 4$ matrix and imposing the rank-one condition for the target state $\alpha\ket{\psi(x,y)} + \beta e^{i\theta(x,y)}\ket{0}$, one obtains a single complex constraint whose modulus reduces to Eq.~(\ref{eq:circle-constraint}). Geometrically, Eq.~(\ref{eq:circle-constraint}) describes the intersection of the Bloch sphere with a plane, i.e., a circle. A finite Kraus decomposition therefore yields a finite union of circles, hence a set of measure zero. A detailed derivation appears in Ref.~\cite{Li2023}.
\end{proof}

{\bf Proposition~\ref{prop:circle-geometry}} has a clear interpretation in our unified framework. A fixed device can only succeed on the states that share a common gauge-fixing relation with respect to the anchor; equivalently, it succeeds only on the families where the anchor-overlap information needed for Eq.~(\ref{eq:norm-overlap}) is already determined. This is precisely the ``known-plus-unknown'' analogue of the unknown-plus-unknown obstruction: one cannot promote an arbitrary unknown ray to a phase-aligned representative without importing the overlap-determinability.

\subsubsection{Probabilistic constructions from partial information as resource conversion} 

The probabilistic protocols that succeed under partial prior information are sometimes presented as exceptions to the no-go theorem. However, in our unified perspective, they also provide an operational demonstration that the overlap-determinability is exactly the missing resource, and they succeed precisely because the missing resource is supplied as a promise.

The canonical example is the protocol of Ref.~\cite{Oszmaniec2016}. Fix a reference state $\ket{\chi}$ and assume the overlaps
\begin{eqnarray}
\tr{\bigl( \ketbra{\chi}{\chi}\hat{\rho}_\psi \bigr)} = c_1 > 0, \quad \tr{\bigl( \ketbra{\chi}{\chi}\hat{\rho}_\phi \bigr)} = c_2 > 0,
\label{eq:fixed-overlaps}
\end{eqnarray}
are known constants. Under Eq.~(\ref{eq:fixed-overlaps}), one can define a coherent superposition that is invariant under independent global-phase changes of
$\ket{\psi}$ and $\ket{\phi}$, namely
\begin{eqnarray}
\label{eq:reference-superposition}
\ket{\Psi_{\alpha,\beta}(\psi,\phi;\chi)} \propto \alpha \frac{\braket{\chi}{\phi}}{\abs{\braket{\chi}{\phi}}}\ket{\psi} + \beta \frac{\braket{\chi}{\psi}}{\abs{\braket{\chi}{\psi}}}\ket{\phi}.
\end{eqnarray}
The phase factors in Eq.~(\ref{eq:reference-superposition}) are a concrete gauge fixing induced by the reference $\ket{\chi}$. The protocol in Ref.~\cite{Oszmaniec2016} constructs an explicit CPTNI map (via a controlled-SWAP and postselection) that outputs Eq.~(\ref{eq:reference-superposition}) with success probability
\begin{eqnarray}
P_{\rm succ} = \frac{c_1c_2}{c_1+c_2}\norm{\ket{\Psi_{\alpha,\beta}(\psi,\phi;\chi)}}^2,
\end{eqnarray}
and subsequent works refined and experimentally implemented the variants~\cite{Dogra2018,Hu2016}.

From our resource viewpoint, the logic is now transparent. The promise in Eq.~(\ref{eq:fixed-overlaps}) supplies precisely the overlap information needed to make the coherence term in Eq.~(\ref{eq:rho-superposed}) well defined at the level of rays. The CPTNI map does not create the overlap-determinability; it consumes the promised overlap data to convert it into an operational superposition. In other words, the known constructions realize the reverse direction of {\bf Theorem~\ref{thm:overlap-resource}} on a restricted domain:
\begin{center}
{\it overlap data (promise) + postselection} $\Longrightarrow$ {\it coherent superposition}.
\end{center}
This is fully consistent with {\bf Theorem~\ref{thm:unified-nosup}}, which forbids the same conversion without the initial overlap resource.

\subsubsection{From universal superposition to no-cloning and no-signaling} 

Finally, we briefly indicate why a universal violation of the no-superposition principle would not remain an isolated oddity. Once a device can superpose generic unknown inputs, {\bf Theorem~\ref{thm:overlap-resource}} upgrades it to an operational overlap oracle, and the overlap information is a powerful currency in quantum information theory.

First, the overlap access collapses the operational gap between linearly dependent and linearly independent families. By composing a hypothetical universal superposer with simple postselection routines, one can map certain linearly dependent input triples to linearly independent output triples, which would enable the unambiguous discrimination and probabilistic cloning on the families that are otherwise forbidden. Second, in entanglement-assisted scenarios, such enhanced discrimination can be turned into the faster-than-light signaling-type contradictions.

The next Sec.~\ref{sec:connections} develops these implications systematically and connects them to standard no-go principles (e.g., no-cloning, no-signaling, and related impossibilities), making explicit how the overlap-determinability resource sits at their common core.

\section{Connections to No-Cloning, No-Signaling, and Other No-Go Theorems}\label{sec:connections}

Here, we complete the narrative that {\bf Theorem~\ref{thm:overlap-resource}} and {\bf Theorem~\ref{thm:unified-nosup}} sit at a common root of several familiar quantum no-go principles. First, we show that a universal probabilistic superposer would convert a linearly dependent family into independent one by attaching a fixed auxiliary direction, thereby enabling the probabilistic cloning beyond the Duan-Guo limit~\cite{Duan1998}. Second, we recall how such enhanced discrimination/cloning can be combined with steering to violate the no-signaling. Finally, we explain how the same overlap-centric obstruction echoes in no-deleting~\cite{Pati2000} and no-masking~\cite{Modi2018}, and why restricted-family (partial-information) constructions remain perfectly consistent: they do not generate the overlap information but consume it.

\subsection{Universal superposition would enable probabilistic cloning beyond linear independence}\label{subsec:superposition-implies-cloning}

We begin by assuming a universal superposition device. Fix nonzero amplitudes $\alpha,\beta \in \mathbb{C}$ with $\abs{\alpha}^2+\abs{\beta}^2=1$. Suppose that there exists a CPTNI map $\Lambda_{\alpha,\beta}$ that is universal in the sense of {\bf Definition~\ref{def:universal_superposer}}. Equivalently, for every pair of pure input rays $\hat{P}_\psi,\hat{P}_\phi$, there exist a phase $\theta(\hat{P}_\psi, \hat{P}_\phi) \in [0,2\pi)$ and a success probability $p_{\psi,\phi}>0$ such that
\begin{eqnarray}
\Lambda_{\alpha,\beta}\Bigl(\hat{P}_\psi \otimes \hat{P}_\phi\Bigr) = p_{\psi,\phi}\ \hat{P}_{\Psi_{\psi,\phi}}, 
\quad
\ket{\Psi_{\psi,\phi}} \propto \alpha\ket{\psi} + \beta e^{i\theta(\hat{P}_\psi,\hat{P}_\phi)}\ket{\phi}.
\label{eq:universal-superposer-assumption}
\end{eqnarray}
This is the direct negation of the no-superposition statement of Ref.~\cite{Oszmaniec2016} and of {\bf Theorem~\ref{thm:unified-nosup}}.

At a conceptual level, the tension is already visible from {\bf Theorem~\ref{thm:overlap-resource}}: if Eq.~(\ref{eq:universal-superposer-assumption}) held on a generic domain, then the associated phase-selection rule would make the overlaps operationally accessible. The overlap access is a powerful currency. In particular, it collapses the usual separation between the linearly dependent and independent families, because the relative-phase data that distinguishes the coherent embeddings in a larger Hilbert-space is precisely the overlap. The next proposition provides a concrete instantiation of this collapse by exhibiting an explicit ``linear-dependence to linear-independence'' reduction.
\begin{proposition}[Universal superposition implies linear-dependence $\rightarrow$ linear-independence]
\label{prop:ld-to-li}
Let $\dim\mathcal{H} \ge 3$ and suppose a universal probabilistic superposition map $\Lambda_{\alpha,\beta}$ satisfying Eq.~(\ref{eq:universal-superposer-assumption}) exists for some fixed nonzero $\alpha,\beta$. Then, there exist three linearly dependent pure states $\{\ket{\psi_1},\ket{\psi_2},\ket{\psi_3}\} \subset \mathcal{H}$ and a fixed auxiliary pure state $\ket{\varphi} \in \mathcal{H}$ such that, conditioned on the success event of $\Lambda_{\alpha,\beta}$ applied to $\ket{\psi_j}\otimes\ket{\varphi}$, the three output states $\{\ket{\Psi_j}\}$ are linearly independent.
\end{proposition}

The proof is given in Appendix~\ref{app:overlap-resource}. Here we highlight the operational meaning. The auxiliary state $\ket{\varphi}$ provides a new orthogonal direction, so the universal output $\alpha\ket{\psi_j} + \beta e^{i\theta_j}\ket{\varphi}$ embeds a two-dimensional (hence linearly dependent) input triple into a three-dimensional subspace. Generically, this embedding produces the linear independence, unless the phases $\{\theta_j\}$ satisfy a special relation. However, the superposition device is a physical map on rays, so the required relation cannot depend on the basis-dependent coordinates of $\ket{\psi_3}$. In the explicit construction in Appendix~\ref{app:proof-overlap-resource}, the ``conspiratorial'' phase constraint is shown to depend on the decomposition parameters of $\ket{\psi_3}$, or equivalently on the overlap data between $\ket{\psi_3}$ and the other inputs. This is exactly where our overlap-determinability enters: preventing the forbidden linear-independence upgrade would require the device to know the overlap information that it is not allowed to manufacture from unknown inputs.

Once {\bf Proposition~\ref{prop:ld-to-li}} is established, the implications for discrimination and cloning follow immediately from standard characterizations.
\begin{corollary}[Universal superposition implies forbidden discrimination and cloning]
\label{cor:superposition-implies-cloning}
Under the assumptions of {\bf Proposition~\ref{prop:ld-to-li}}, there exists a probabilistic protocol that
\begin{itemize}
\item[\emph{(i)}] unambiguously discriminates a linearly dependent set of pure states with nonzero success probability, and hence
\item[\emph{(ii)}] probabilistically clones that linearly dependent set with nonzero success probability.
\end{itemize}
\end{corollary}

Indeed, since the outputs $\{\ket{\Psi_j}\}$ are linearly independent, unambiguous state discrimination exists for them by the Chefles criterion~\cite{Chefles1998,Cai2024}. Conditioned on a successful discrimination outcome, one can reprepare $\ket{\psi_j}^{\otimes 2}$, yielding a probabilistic cloner for a linearly dependent set. This contradicts the Duan-Guo characterization that probabilistic cloning of pure states is possible if and only if the set is linearly independent~\cite{Duan1998}.

It is worth emphasizing how this fits our unified viewpoint. The traditional no-cloning theorem is often presented as a consequence of linearity. Here the contradiction is sharper: a universal superposer would effectively supply the missing phase/overlap data needed to turn linearity itself into an advantage, allowing one to move a dependent family into a regime where discrimination and cloning become possible. In this sense the no-superposition principle is not merely parallel to no-cloning; it sits upstream, because it forbids the overlap oracle that would trivialize the linear-independence barrier.

This also clarifies, again, why restricted-state superposition protocols are consistent with the no-go landscape~\cite{Oszmaniec2016,Dogra2018,Hu2016}. Those protocols come with a promise that fixes overlaps with a reference state $\ket{\chi}$, or restrict inputs to geometrically thin subsets (such as circles on the Bloch sphere). In our language, the promise supplies the overlap-determinability resource externally. The protocol does not violate no-cloning because it never upgrades linear dependence without paying for the missing overlap information.

\subsection{No-signaling: why a universal superposer would be unphysical even beyond quantum theory}\label{subsec:superposition-implies-nosignaling}

The previous subsection already shows that the universal superposition would collapse the operational separation between ``allowed'' and ``forbidden'' discrimination/cloning tasks. A natural next question is whether this collapse is merely a peculiarity of quantum mechanics, or whether it is incompatible with more primitive principles. The answer is that it is incompatible even with no-signaling.

The no-signaling principle asserts that the local actions on Alice's system cannot influence the statistics of measurements on Bob's system at space-like separation. In standard quantum theory, this is guaranteed by the complete positivity and the tensor-product structure. However, once one admits a device that enables the cloning or unambiguous discrimination beyond linear independence, the steering scenarios become signaling machines: Alice can remotely prepare different ensemble decompositions of the same reduced state on Bob's side, and Bob can exploit the forbidden primitive to distinguish which decomposition he received.

The following proposition formalizes the key step.
\begin{proposition}[Probabilistic cloning of a linearly dependent set violates no-signaling]
\label{prop:probcloning-violates-nosignaling}
Suppose there exists a probabilistic protocol that clones each state in a known linearly dependent set $\mathcal{S}=\{\ket{\psi_j}\}_{j=1}^n$ with nonzero success probability. Then, there exists a bipartite steering scenario in which Alice can choose between two local measurements that prepare two distinct ensemble decompositions of the same reduced state on Bob's side, and Bob can exploit the cloning protocol to distinguish Alice's choice with nonzero advantage. Repetition amplifies this advantage to enable super-luminal signaling.
\end{proposition}

\begin{proof}---Assume Bob possesses a probabilistic cloner for a known linearly dependent set $\mathcal{S}=\{\ket{\psi_j}\}_{j=1}^n$ on $\mathcal{H}_B$. Concretely, there exists a physical instrument with a ``success'' outcome such that, for every $j$,
\begin{eqnarray}
\ket{\psi_j}\ket{0}\ \mapsto\ \ket{\psi_j}\ket{\psi_j}
\end{eqnarray}
with some success probability $p_j>0$. Then, let $\hat{\rho}_B$ be any mixed state that admits two distinct ensemble decompositions,
\begin{eqnarray}
\hat{\rho}_B = \sum_{j=1}^n q_j \ket{\psi_j}\bra{\psi_j} = \sum_{k} r_k \ket{\phi_k}\bra{\phi_k},
\label{eq:two-decompositions}
\end{eqnarray}
with $\{\ket{\psi_j}\}\subseteq\mathcal{S}$. Such alternative decompositions always exist for a nontrivial $\hat{\rho}_B$. By the Hughston-Jozsa-Wootters theorem~\cite{Hughston1993}, there exists a purification $\ket{\Omega}_{AB}$ of $\hat{\rho}_B$ and two different measurements $\hat{M}_0$ and $\hat{M}_1$ on Alice's system such that: (i) if Alice performs $\hat{M}_0$, then Bob's post-measurement state is $\ket{\psi_j}$ with probability $q_j$; (ii) if Alice performs $\hat{M}_1$, then Bob's post-measurement state is $\ket{\phi_k}$ with probability $r_k$. In either case, Bob's unconditional reduced state remains $\hat{\rho}_B$; hence, if Bob is restricted to ordinary quantum operations, he cannot learn which measurement Alice performed.

Now suppose Alice uses her measurement choice as a one-bit message: $\hat{M}_0$ encodes ``0'' and $\hat{M}_1$ encodes ``1''. Bob attempts to decode as follows. On each run he feeds his received state together with an ancilla $\ket{0}$ into his probabilistic cloner and checks whether the success flag clicks. When Alice chooses $\hat{M}_0$, the input is guaranteed to be one of the $\ket{\psi_j} \in \mathcal{S}$, hence the success event occurs with probability
\begin{eqnarray}
P_{\mathrm{succ}|0} = \sum_{j=1}^n q_j p_j > 0.
\label{eq:succ0}
\end{eqnarray}
When Alice chooses $\mathsf{M}_1$, Bob receives states $\ket{\phi_k}$ that, in general, need not belong to $\mathcal{S}$. Even if the cloner occasionally succeeds on some of them, the overall success probability
\begin{eqnarray}
P_{\mathrm{succ}|1} = \sum_{k} r_k p(\phi_k)
\label{eq:succ1}
\end{eqnarray}
will typically differ from Eq.~(\ref{eq:succ0}) unless the device is conspiratorially tuned. In particular, since $\mathcal{S}$ is linearly dependent and finite, one can always pick a decomposition in Eq.~(\ref{eq:two-decompositions}) such that at least one $\ket{\phi_k}$ lies outside the support on which the probabilistic cloner succeeds (this is the content of the no-signaling constraints analyzed in Refs.~\cite{Gisin1998,Hardy1999}). Hence, $P_{\mathrm{succ}|0} \neq P_{\mathrm{succ}|1}$ for some valid steering scenario.

Bob can therefore distinguish Alice's message with nonzero advantage from the observed success statistics, and by repeating the protocol over many independent shared pairs he amplifies this advantage to arbitrary reliability, yielding super-luminal signaling. This contradicts no-signaling.
\end{proof}

This statement refines a standard theme in the foundations of quantum information: forbidden transformations tend to become signaling machines when combined with entanglement. The specific relation between no-signaling and probabilistic cloning was analyzed by Hardy and Song~\cite{Hardy1999} and discussed in the present context in Ref.~\cite{Bandyopadhyay2020}.

Combining {\bf Proposition~\ref{prop:probcloning-violates-nosignaling}} with {\bf Corollary~\ref{cor:superposition-implies-cloning}}, we obtain a sharp implication: a universal probabilistic superposer of the form Eq.~(\ref{eq:universal-superposer-assumption}) is incompatible with no-signaling. Importantly, this conclusion does not rely on the full structure of quantum mechanics. It only uses (i) steering (or more generally, preparation of different ensembles for the same mixed state) and (ii) cloning/discrimination beyond linear independence. Thus, even within a broader class of no-signaling theories, a universal superposition device cannot be admitted.

From our overlap-determinability viewpoint, this is almost inevitable. A universal coherent superposer would implicitly act as an ``overlap-handling oracle'': it would have to implement a consistent choice of relative phase (hence a consistent overlap relation) that is not recoverable from local density operators. But different ensemble decompositions of the same mixed state are locally indistinguishable precisely because they hide such overlap data. An overlap oracle therefore breaks the operational equivalence that protects no-signaling.

\subsection{Further connections: no-deleting, no-masking, and overlap reshaping}\label{subsec:other-nogos}

The arguments above show that if one violates the no-superposition principle universally, the violation propagates through the standard no-go architecture: the universal superposition would enable forbidden discrimination/cloning and would thereby violate no-signaling. It is natural to ask how far this propagation extends into other no-go theorems. Here, we briefly comment on two further links that are particularly transparent from the overlap viewpoint.

\subsubsection{No-deleting as the time-reverse shadow of cloning}  

The no-deleting theorem can be viewed as a conceptual dual of no-cloning~\cite{Pati2000,Pati2003}. Its overlap content is immediate. Suppose a reversible (isometric) transformation $\hat{V}$ could delete one copy universally,
\begin{eqnarray}
\hat{V}\ket{\psi}\ket{\psi} = \ket{\psi}\ket{\mathbb{0}} \quad (\forall \, \ket{\psi}).
\label{eq:delete_map}
\end{eqnarray}
By preservation of inner products under isometries, we would have, for all $\ket{\psi},\ket{\phi}$,
\begin{eqnarray}
\braket{\psi}{\phi}^2 = \braket{\psi, \psi}{\phi, \phi} = \braket{\psi, \mathbb{0}}{\phi, \mathbb{0}}
= \braket{\psi}{\phi}.
\label{eq:delete_overlap}
\end{eqnarray}
Thus, $\braket{\psi}{\phi}$ would have to satisfy $x^2=x$ for all possible overlaps, which forces $x \in \{0,1\}$. The universal deleting (like universal cloning) is therefore impossible except on orthogonal/identical families. In our language, both tasks are instances of forbidden overlap reshaping: they demand a nontrivial functional transformation of the overlap data that cannot arise from physical linear dynamics.

\subsubsection{No-masking and the geometry of restricted families.}   

The no-masking theorem states tells us the following~\cite{Modi2018,Li2020,Lie2020}: there is no universal isometry that maps an arbitrary unknown pure state to a bipartite state whose reduced states are independent of the input, i.e., there is no $\hat{V}$ such that for all $\ket{\psi}$,
\begin{eqnarray}
\ket{\psi}\ \mapsto\ \ket{\Psi_\psi}_{AB},
\quad
\text{Tr}_{B}{\Bigl(\ket{\Psi_\psi}\bra{\Psi_\psi}\Bigr)}=\hat{\sigma}_A,
\quad
\text{Tr}_{A}{\Bigl(\ket{\Psi_\psi}\bra{\Psi_\psi}\Bigr)}=\hat{\sigma}_B,
\label{eq:masking-def}
\end{eqnarray}
with fixed marginals $\hat{\sigma}_A, \hat{\sigma}_B$ independent of $\psi$. Because $\hat{V}$ is an isometry, it must preserve inner products:
\begin{eqnarray}
\braket{\psi}{\phi} = \braket{\Psi_\psi}{\Psi_\phi} \quad (\forall \, \ket{\psi},\ket{\phi}).
\label{eq:masking_overlap}
\end{eqnarray}
Thus, all distinguishability information (the Gram-matrix structure of the input family) must be carried entirely by correlations while the marginals remain frozen. For arbitrary inputs, this requirement is too rigid. What matters for the present paper is the structural parallel: no-masking, like no-superposition, admits nontrivial restricted-family constructions, and those constructions are possible precisely when the overlap/phase degrees of freedom are constrained in advance. For qubits, families on a circle (e.g., equatorial phase families) can be masked by embedding the unknown relative phase into correlations, just as families with fixed overlaps to a reference can be coherently superposed~\cite{Oszmaniec2016,Dogra2018,Li2023}. In both cases, the permitted families are ``thin'' because the missing resource is the same: one is not allowed to manufacture overlap/phase information universally from independent unknown preparations.

The upshot is that our main theorem should be read as more than a statement about superpositions. It identifies a single structural obstruction---universal access to overlap information---whose violation would not only create ``superpositions'' but would also unravel the broader web of no-go principles that protect quantum theory (and, more generally, no-signaling theories) from classical-style information processing. In this sense, overlap-determinability is a unifying currency: it explains at once why superposition can be engineered on restricted families, why the universal superposition fails, and why the hypothetical universal superposition would imply cloning and signaling contradictions.

\section{Arbitrary superposition and the collapse of Grover's lower bound}\label{sec:grover}

Quantum mechanics famously grants an exponential state space while withholding an exponential generic advantage for unstructured search. In the standard black-box model, one is given oracle access to a Boolean function $f: \{0,1\}^{n} \rightarrow \{0,1\}$ promised to have a unique marked input $w$ such that $f(w)=1$, and the goal is to identify $w$. The oracle can be taken, without loss of generality, to be the phase oracle
\begin{eqnarray}
\hat{O}_{f}\ket{x}=(-1)^{f(x)}\ket{x}.
\label{eq:grover-oracle}
\end{eqnarray}
Grover's algorithm finds $w$ using $O(\sqrt{N})$ oracle calls, where $N:=2^{n}$~\cite{Grover1997}, and the Bennett-Bernstein-Brassard-Vazirani hybrid argument shows that this scaling is optimal for quantum theory: i.e., any algorithm using unitary dynamics and Born-rule measurement requires $\Omega(\sqrt{N})$ queries~\cite{Bennett1997} (see also Ref.~\cite{Zalka1999}). Even though the computation can explore $N$ basis states coherently, the theory enforces a sharp boundary.

In this section, we connect this algorithmic boundary to our central resource viewpoint. We take the counterfactual possibility that arbitrary unknown-state superposition is available as a primitive operation. In our language, this corresponds to promoting the overlap-determinability to a free resource on a generic domain, violating {\bf Theorem~\ref{thm:overlap-resource}} and hence the unified no-superposition principle of {\bf Theorem~\ref{thm:unified-nosup}}. We will argue that the consequences would be dramatic not only for quantum information tasks but also for computational complexity: the Grover lower bound would collapse, enabling exponential speedups.

The bridge is a striking observation due to Bao, Bouland, and Jordan~\cite{Bao2016}: within several broad and physically motivated classes of the post-quantum deviations---including nonunitary maps, modified Born rules, cloning, and postselection---the ability to send super-luminal signals is equivalent to the ability to beat Grover's lower bound, and the physical resources needed for one scale polynomially with the resources needed for the other. Since Sec.~\ref{sec:connections} already showed that a universal superposition device would imply super-luminal signaling, Bao \emph{et al.}'s equivalence immediately upgrades our overlap-determinability obstruction into a complexity-theoretic one.

\subsection{Grover search, overlap geometry, and what the lower bound is really constraining}\label{subsec:grover-overlap-geometry}

It is useful to recall that Grover's iterate is, geometrically, a sequence of reflections in a two-dimensional plane. Let
\begin{eqnarray}
\ket{\psi_0} := \frac{1}{\sqrt{N}}\sum_{x=0}^{N-1}\ket{x}
\label{eq:uniform-superposition}
\end{eqnarray}
be the uniform state and $\ket{w}$ the marked basis state. The usual Grover operator can be written as
\begin{eqnarray}
\hat{G} := \left(2\ketbra{\psi_0}{\psi_0}-\hat{\mathds{1}}\right)\hat{O}_{f},
\label{eq:grover-iterate}
\end{eqnarray}
and its action is a rotation in the plane spanned by $\ket{\psi_0}$ and $\ket{w}$~\cite{Grover1997,Zalka1999}. Here, the explicit form of the oracle operation is $\hat{O}_f = 2\ketbra{w}{w} - \hat{\mathds{1}}$. The angle of rotation is controlled by the overlap $\abs{\braket{\psi_0}{w}}=1/\sqrt{N}$, which is exponentially small in the number of qubits. This is the quantitative reason for the $\sqrt{N}$ scaling: each iteration rotates by an angle $\Theta(1/\sqrt{N})$.

From the perspective of the present study, this is already a familiar pattern: an algorithmic speed limit is dictated by an overlap term, and the relevant overlap is not an arbitrary complex number but a physically meaningful phase-fixed quantity. In ordinary Grover search, this phase structure is benign because $\ket{\psi_0}$ is ``known'' (it is prepared by a known circuit) and $\ket{w}$ is a computational basis state. However, if one enriches the theory with an operation that can coherently combine ``unknown'' inputs while fixing their relative phase---that is, if one makes the overlap-determinability free on generic domains---one is precisely endowing the experimenter with a new ability to manufacture and control interference terms that are not determined by density operators alone. In Sec.~\ref{sec:overlap}, we identified this as the missing resource behind the universal superposition. Here, we emphasize that the same resource, if available universally, would allow one to engineer rotations in exponentially small-overlap regimes far more aggressively than unitary Grover dynamics permits.

\subsection{Bao--Bouland--Jordan: no-signaling and Grover are two sides of the same coin} \label{subsec:bao-equivalence}

Bao \emph{et al.}~\cite{Bao2016} investigated a question that looks, at first glance, orthogonal to ours: why does the exponential Hilbert space not trivialize NP-hard problems by brute-force search? Their answer is that, within broad classes of post-quantum models, the obstruction that prevents faster-than-light signaling is quantitatively tied to the obstruction that prevents speedups over Grover search.

Concretely, Ref.~\cite{Bao2016} considers four classes of deviations from standard quantum theory, motivated in part by black-hole information discussions: (i) nonunitary but linear maps (final-state projection type dynamics), (ii) modifications of the Born rule, (iii) the addition of a cloning primitive, and (iv) the addition of a postselection primitive. In each case, they prove an ``if and only if'' statement: the model admits super-luminal signaling if and only if it admits a query-complexity speedup over Grover search. Moreover, the quantitative tradeoff is polynomial:
if one can transmit even a single classical bit super-luminally using $m$ qubits and $\mathrm{poly}(m)$ operations, then one can speed up Grover search on an exponentially larger instance size $N=2^{\mathrm{poly}(m)}$ with $\mathrm{poly}(m)$ resources, and conversely. In short, within these models,
\begin{eqnarray}
\mbox{``reasonable-resource signaling''}\ \Longleftrightarrow\ \mbox{``reasonable-resource exponential search speedups.''}
\label{eq:bao-slogan}
\end{eqnarray}

Bao \emph{et al.} do not require the full structure of quantum mechanics to derive their equivalence. Rather, the proofs exploit the same structural ingredients that appear repeatedly in our analysis: nonlinearity on states induced by postselection-like primitives, the operational distinguishability of different ensemble decompositions of the same density operator, and the amplification of exponentially small distinguishability gaps by iterating a nonlinear transformation. These are precisely the patterns that a universal superposition device would enable, because it would operationalize the overlap data that is otherwise hidden by the ray gauge freedom.

\subsection{From overlap-determinability violation to exponential Grover speedups}\label{subsec:superposition-to-grover}

We can now state the computational consequence of our no-superposition principle as a sharp corollary.
\begin{corollary}[Universal superposition would imply exponential speedups over Grover search]
\label{cor:superposition-implies-grover}
Assume, contrary to {\bf Theorem~\ref{thm:unified-nosup}}, that there exists a universal probabilistic superposition device: a physical primitive that, on generic unknown inputs, outputs a coherent state proportional to $\alpha\ket{\psi} + \beta e^{i\theta(\hat{P}_\psi,\hat{P}_\phi)}\ket{\phi}$ with nonzero success probability. Then, the resulting theory admits super-luminal signaling with finite resources, and consequently admits a query-complexity speedup over Grover search. In particular, by the equivalence of Ref.~\cite{Bao2016} (in the cloning/postselection classes), one obtains an exponential speedup: there exist unstructured search instances of size $N$ that can be solved with $\mathrm{poly}(\log{N})$ queries and $\mathrm{poly}(\log{N})$ additional operations.
\end{corollary}

\begin{proof}---The first ingredient is already established in Sec.~\ref{sec:connections}: postulating a universal coherent superposer endows the theory with operational access to phase-sensitive overlap information from independently prepared unknown rays, i.e., it entails an overlap-determinability violation and hence a usable ``overlap oracle.''

A particularly transparent route from such overlap control to a Grover collapse is through unknown-state reflections~\cite{Kumar2011}. Given an ``a priori unknown'' pure state $\ket{\psi}$, consider the reflection
\begin{eqnarray}
\hat{R}_{\psi} := 2\ketbra{\psi}{\psi} - \hat{\mathds{1}}.
\end{eqnarray}
If such a reflection $\hat{R}_{\psi}$ is available at nonzero constant cost, then Grover's lower bound would collapse.
To see this directly, define a variant of Grover iteration by setting $\ket{\psi_0}$ as in Eq.~(\ref{eq:uniform-superposition})
and updating, for $r=0,1,2,\dots$,
\begin{eqnarray}
\ket{\psi_{r+1}} := \hat{R}_{\psi_r} \hat{O}_{f}\,\ket{\psi_r}.
\label{eq:previous-output-iterate}
\end{eqnarray}
Each round uses exactly one oracle call $\hat{O}_f$, but the reflection axis of $\hat{R}_{\psi_r}$ is updated from the fixed $\ket{\psi_0}$ to the current (hence unknown)
state $\ket{\psi_r}$.

Let $a_r:=\braket{w}{\psi_r}$ denote the marked-state overlap amplitude. A short calculation shows that this overlap obeys a nonlinear recursion that amplifies exponentially for small $|a_r|$:
\begin{eqnarray}
a_{r+1} &=& \bra{w}\hat{R}_{\psi_r}\hat{O}_f\ket{\psi_r} = 2\braket{w}{\psi_r}\bra{\psi_r}\hat{O}_f\ket{\psi_r} - \bra{w}\hat{O}_f\ket{\psi_r} \nonumber\\
	&=& 2a_r \bigl( 1-2 \abs{a_r}^2 \bigr) - (-a_r) = \bigl( 3 - 4\abs{a_r}^2 \bigr)a_r.
\label{eq:overlap-recursion}
\end{eqnarray}
Equivalently, the success probability $p_r := \abs{a_r}^2$ satisfies $p_{r+1}=p_r(3-4p_r)^2$. In particular, as long as $p_r\le 1/4$ we have $(3-4p_r)\ge 2$, hence
\begin{eqnarray}
p_{r+1} \ge 4p_r.
\label{eq:prob-growth}
\end{eqnarray}
Since $p_0=\abs{\braket{w}{\psi_0}}^2=1/N$, it follows inductively that $p_r \ge 4^r/N$ until $p_r$ reaches a constant. Choosing
\begin{eqnarray}
r(N)=\left\lceil \log_4{\left(\frac{N}{4}\right)} \right\rceil = \Theta(\log N)
\label{eq:log-rounds}
\end{eqnarray}
ensures $p_{r(N)}\ge 1/4$, i.e., a constant success bias after only $\Theta(\log N)$ oracle queries. Repeating a constant number of times boosts the success probability arbitrarily close to $1$. Thus, in the unit-cost model for $\hat{R}_{\psi_r}$, the $\Theta(\sqrt{N})$ Grover lower bound collapses to $\mathrm{poly}(\log N)$ query complexity.

This also makes clear why such reflections are precisely where the overlap-determinability obstruction bites. Even on a known test state $\ket{\phi}$, the action of an unknown-state reflection necessarily manufactures an interference term whose coefficient is a
phase-sensitive overlap:
\begin{eqnarray}
\hat{R}_{\psi_r}\ket{\phi} = \left(2\ketbra{\psi_r}{\psi_r} - \hat{\mathds{1}}\right)\ket{\phi} = 2\braket{\psi_r}{\phi}\ket{\psi_r} - \ket{\phi}.
\label{eq:reflection-requires-overlap}
\end{eqnarray}
When $\ket{\psi_r}$ is oracle-dependent and not classically specified, the complex number $\braket{\psi_r}{\phi}$ is exactly the kind of convention-fixed overlap that is not determined by the rays alone. Granting $\hat{R}_{\psi_r}$ as a constant-cost primitive therefore amounts to granting, in operational form, access to the missing overlap/phase information that our no-superposition principle forbids from being generated universally.

Finally, by Sec.~\ref{sec:connections} this overlap-determinability violation already implies super-luminal signaling with finite resources. Equivalently, the Bao--Bouland--Jordan equivalence converts any such super-Grover speedup into a super-luminal classical channel with only polynomial overhead. This completes the proof.
\end{proof}

{\bf Corollary~\ref{cor:superposition-implies-grover}} is a concrete algorithmic avatar of the no-universal superposition. Nevertheless, if one could control overlaps well enough to implement the unknown-state reflections on demand, then the dynamics would become available and would immediately yield the logarithmic-round (``super-Grover'') amplification, collapsing Grover's lower bound~\cite{Yee2020}. In this sense, the no-superposition principle can be read as protecting both relativistic causality and the computational lower bounds enforced by it (see also Ref.~[Bang:2026]).

\section{Discussion and outlook}

We have argued that the obstacle behind universal superposition is not merely probabilistic, but fundamentally phaselike: a ray specifies only a projector, while coherent interference requires a definite complex overlap between chosen representatives. Concretely, if the inputs are $\hat{\rho}_{\psi}=\ketbra{\psi}{\psi}$ and $\hat{\rho}_{\phi}=\ketbra{\phi}{\phi}$, then any CPTNI processing is, by construction, a function of these projectors. However, a target ray proportional to $\alpha\ket{\psi}+\beta e^{i\theta}\ket{\phi}$ necessarily contains the interference terms of the form $\alpha\beta^{*}e^{-i\theta}\ketbra{\psi}{\phi}+\mathrm{H.c.}$, whose value depends on a phase choice for the input representatives. Indeed, the independent rephasings $\ket{\psi} \mapsto e^{i\varphi}\ket{\psi}$ and $\ket{\phi} \mapsto e^{i\varphi'}\ket{\phi}$ leave $\hat{\rho}_{\psi},\hat{\rho}_{\phi}$ invariant but rotate $\ketbra{\psi}{\phi}$ by $e^{i(\varphi-\varphi')}$, so the cross term that drives the interference is not determined by density operators alone. Thus a ``universal superposer'' would have to do more than occasionally succeed: it would have to supply an input-independent rule that turns the rays into phased vectors, thereby fixing an otherwise gauge-dependent interference term. To isolate exactly what is missing, we introduced phase conventions and the notion of ``overlap-determinability'' ({\bf Definition~\ref{def:overlapdet}}). A phase convention $\Gamma$ is precisely such a lifting: it assigns to each ray $\hat{\rho} \in \mathcal{R}$ a normalized representative vector; writing $\ket{\psi_{\Gamma}}:=\Gamma(\hat{\rho}_{\psi})$ and $\ket{\phi_{\Gamma}}:=\Gamma(\hat{\rho}_{\phi})$, the convention-fixed overlap $\braket{\psi_{\Gamma}}{\phi_{\Gamma}}$ becomes a well-defined operational datum, and the interference operator (and hence the superposed output ray) becomes unambiguous. Our main result, {\bf Theorem~\ref{thm:unified-nosup}}, then unifies modern no-superposition theorems by showing that a probabilistic superposition map exists on a domain if and only if that domain is overlap-determinable. In particular, for generic independently prepared unknown rays, no CP dynamics can manufacture such an input-independent lifting, so such a map cannot exist. Beyond the superposition task itself, Secs.~\ref{sec:connections} and \ref{sec:grover} showed that treating convention-fixed overlaps as freely obtainable immediately destabilizes both the causal structure (via steering-to-signaling) and the usual computational geometry of amplitude amplification.

\begin{figure}[t]
\centering
\includegraphics[width=1.00\linewidth]{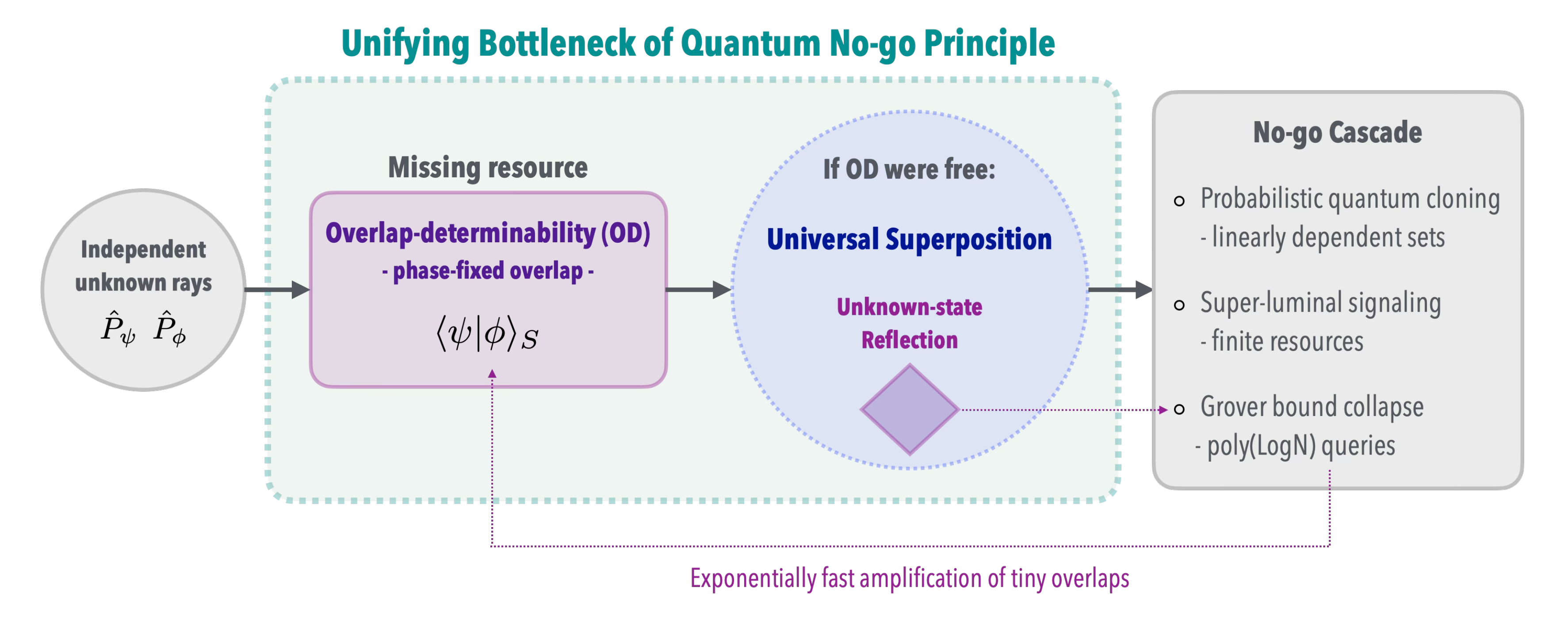}
\caption{\textbf{Overlap-determinability as a unifying bottleneck.} Independently prepared unknown rays do not, by themselves, determine a phase-fixed overlap. If a scenario were to make overlap-determinability freely available on generic domains, it would enable primitives such as universal superposition and unknown-state reflections; these, in turn, trigger a cascade of post-quantum consequences, including probabilistic cloning of linearly dependent families, finite-resource superluminal signaling, and the collapse of Grover's query lower bound via aggressive overlap amplification.}
\label{fig:od-roadmap}
\end{figure}

From a broader perspective, our overlap-determinability functions as a phase reference for rays. When it is supplied by a physical scenario (promises, reference systems, or classical side information), it licenses the operations whose behavior depends explicitly on interference terms such as $\ketbra{\psi}{\phi}$---objects that are not determined by the input density operators alone. This single new license ties together a constellation of no-go statements that are usually discussed separately. If the overlap-determinability is available generically at constant cost, one could (i) coherently superpose unknown inputs with a well-defined relative phase, and (ii) implement reflections about oracle-dependent intermediate states, which amounts to amplifying an initially tiny marked-state overlap far more aggressively than Grover's unitary rotation permits. In both routes, the story is the same: access to phase-fixed overlaps turns exponentially small geometric angles into controllable dynamical levers, and the familiar barriers---no-cloning/no-signaling on the foundational side and Grover's lower bound on the algorithmic side---collapse together. We provide a conceptual map of these implications in Fig.~\ref{fig:od-roadmap}.

Several open problems are suggested by this viewpoint. First, our results are largely qualitative; it is natural to develop quantitative notions of approximate overlap-determinability and relate the achievable error and success probability to operational resources (number of copies, reference frames, gate complexity, etc.). Second, many realistic settings provide partial phase information (e.g., known overlaps with a reference state, symmetry-restricted families, or structured manifolds such as Gaussian states); characterizing the maximal overlap-determinable domains under such promises could turn the present no-go principle into a practical design rule for ``when superposition is possible.'' Third, on the computational side, it would be valuable to chart intermediate models between standard Grover dynamics and an ideal unknown-reflection oracle: how accurately must one approximate a reflection about an unknown intermediate state to obtain any super-Grover advantage, and what are the corresponding lower bounds on learning/control complexity? Finally, it remains intriguing to ask whether overlap-determinability admits a natural resource-theoretic formulation, placing superposition, reflection, and signaling phenomena on a common quantitative axis, and whether experimentally accessible partial instances can be harnessed without crossing into causal or complexity-theoretic pathologies.

\section*{Acknowledgement}
This work was supported by the Ministry of Science, ICT and Future Planning (MSIP) by the National Research Foundation of Korea (RS-2024-00432214, RS-2025-03532992, and RS-2025-18362970) and the Institute of Information and Communications Technology Planning and Evaluation grant funded by the Korean government (RS-2019-II190003, ``Research and Development of Core Technologies for Programming, Running, Implementing and Validating of Fault-Tolerant Quantum Computing System''), the Korean ARPA-H Project through the Korea Health Industry Development Institute (KHIDI), funded by the Ministry of Health \& Welfare, Republic of Korea (RS-2025-25456722). We acknowledge the Yonsei University Quantum Computing Project Group for providing support and access to the Quantum System One (Eagle Processor), which is operated at Yonsei University.

\appendix

\section{Full proof of Theorem~\ref{thm:overlap-resource}}\label{app:overlap-resource}

In this Appendix, we provide a complete and self-contained proof of {\bf Theorem~\ref{thm:overlap-resource}}. The goal is to make precise (and prove) the core mechanism behind the ``overlap-determinability'' principle, emphasized in Sec.~\ref{sec:overlap}: a nontrivial superposition map must implicitly fix a relative phase convention between two rays, and this is impossible unless some additional overlap/phase reference is available. Our proof is built around a linear-independence obstruction that converts a hypothetical superposition machine into an unambiguous discrimination machine, thereby colliding with the most basic linear-algebraic constraints of quantum theory.

\subsection{Setup: probabilistic superposition maps and phase-gauge covariance}\label{app:setup-superposition}

Let ${\cal H}$ be a complex Hilbert space with $\dim{\cal H} \ge 2$. For a unit vector $\ket{\psi}\in{\cal H}$ we denote the corresponding pure state (ray) by
\begin{eqnarray}
\hat{P}_\psi := \ketbra{\psi}{\psi}.
\end{eqnarray}
A CPTNI map ${\cal E}$ admits a Kraus representation
\begin{eqnarray}
{\cal E}(\hat{\rho}) = \sum_k \hat{M}_k \hat{\rho} \hat{M}_k^\dagger, \quad \sum_k \hat{M}_k^\dagger \hat{M}_k \le I,
\label{eq:app-kraus}
\end{eqnarray}
and its success probability on input $\hat{\rho}$ is $\tr{{\cal E}(\hat{\rho})}$.

For fixed nonzero complex numbers $\alpha,\beta \neq 0$ (with $\abs{\alpha}^2 + \abs{\beta}^2=1$, we adopt the operational notion of superposition for rays (rather than vectors) in the sense of Refs.~\cite{Oszmaniec2016,Dogra2018,Hu2016,Li2017}. A probabilistic superposition map for weights $(\alpha,\beta)$ is a CPTNI map
\begin{eqnarray}
\Lambda_{\alpha,\beta} : {\cal B}({\cal H}\otimes{\cal H}) \rightarrow {\cal B}({\cal H})
\end{eqnarray}
such that, for every allowed input pair $(\hat{P}_\psi, \hat{P}_\phi)$ in its domain, the output upon success is a pure state proportional to a coherent two-branch superposition of the two inputs:
\begin{eqnarray}
\Lambda_{\alpha,\beta}(\hat{P}_\psi\otimes \hat{P}_\phi) = p_{\psi,\phi}\,\hat{P}_{\Psi(\psi,\phi)},
\quad
p_{\psi,\phi}>0,
\label{eq:app-superposer-def}
\end{eqnarray}
where $\ket{\Psi(\psi,\phi)}$ is a normalized vector of the form
\begin{eqnarray}
\ket{\Psi(\psi,\phi)} \propto \alpha \ket{\psi} + \beta e^{i\theta(\psi,\phi)} \ket{\phi}
\label{eq:app-superposition-vector-form}
\end{eqnarray}
for some phase $\theta(\psi,\phi)\in[0,2\pi)$ that may depend on the rays $\hat{P}_\psi, \hat{P}_\phi$ but must not depend on unphysical choices of vector representatives. The latter constraint is the phase-gauge covariance that is forced by the fact that a physical operation acts on density operators. Concretely, since $\hat{P}_{e^{i\gamma}\psi}=\hat{P}_\psi$, the ray $\hat{P}_{\Psi(\psi,\phi)}$ produced by $\Lambda_{\alpha,\beta}$ must be invariant under independent rephasings
\begin{eqnarray}
\ket{\psi}\rightarrow e^{i\gamma}\ket{\psi}, \quad \ket{\phi}\rightarrow e^{i\delta}\ket{\phi},
\label{eq:app-gauge}
\end{eqnarray}
in the sense that the output projector $\hat{P}_{\Psi(\psi,\phi)}$ is unchanged. This is exactly the place where ``overlap-determinability'' enters: without some additional phase reference that correlates the gauge in Eq.~(\ref{eq:app-gauge}) between different rays, the relative phase between two unknown rays is not a physical datum that $\Lambda_{\alpha,\beta}$ can access.

\subsection{A lemma: probabilistic CP maps cannot enable unambiguous discrimination of linearly dependent sets}\label{app:lemma-UD}

We first prove a lemma that isolates the key obstruction. It is well known that a finite set of pure states is unambiguously distinguishable (with nonzero total success probability) if and only if the states are linearly independent. We include a short self-contained proof, since it is the logical hinge of {\bf Theorem~\ref{thm:overlap-resource}}.

\begin{lemma}[Unambiguous discrimination and linear independence]
\label{lem:UD-linear-independence}
Let $\{\ket{\varphi_j}\}_{j=1}^m \subset {\cal H}$ be a finite set of unit vectors. There exists a POVM $\{\hat{E}_1,\ldots,\hat{E}_m,\hat{E}_\kappa\}$ such that
\begin{eqnarray}
\bra{\varphi_j}\hat{E}_i\ket{\varphi_j} = 0 \ \ \ (j \neq i \in [1,m]),
\quad
\bra{\varphi_i}\hat{E}_i\ket{\varphi_i} > 0 \ \ \ (\forall \, i \in [1,m]),
\label{eq:app-UD-conditions}
\end{eqnarray}
if and only if the vectors $\{\ket{\varphi_j}\}_{j=1}^m$ are linearly independent.
\end{lemma}

\begin{proof}---Assume first that $\{\ket{\varphi_j}\}_{j=1}^m$ are linearly dependent. Then, there exists an index, say $j=1$, and complex coefficients $c_2,\ldots,c_m$ such that
\begin{eqnarray}
\ket{\varphi_1} = \sum_{j=2}^m c_j \ket{\varphi_j}.
\label{eq:app-linear-dep}
\end{eqnarray}
Suppose a POVM satisfying Eq.~(\ref{eq:app-UD-conditions}) exists. For $i=1$, the condition $\bra{\varphi_j}\hat{E}_1\ket{\varphi_j}=0$ for all $j \neq 1$ implies that $\hat{E}_1$ annihilates the entire span of $\{\ket{\varphi_j}\}_{j=2}^m$, hence by Eq.~(\ref{eq:app-linear-dep}) it also annihilates $\ket{\varphi_1}$. Therefore, $\bra{\varphi_1}\hat{E}_1\ket{\varphi_1}=0$, contradicting the strict positivity in Eq.~(\ref{eq:app-UD-conditions}). Thus, no such POVM exists when the states are linearly dependent.

Conversely, assume that $\{\ket{\varphi_j}\}_{j=1}^m$ are linearly independent. Then, there exists a reciprocal (biorthogonal) family $\{\ket{\widetilde{\varphi}_j}\}_{j=1}^m$ satisfying
\begin{eqnarray}
\braket{\widetilde{\varphi}_i}{\varphi_j} = \delta_{ij}.
\label{eq:app-reciprocal}
\end{eqnarray}
Define POVM elements
\begin{eqnarray}
\hat{E}_i = \lambda_i \ketbra{\widetilde{\varphi}_i}{\widetilde{\varphi}_i},
\quad
\hat{E}_\kappa = I - \sum_{i=1}^m \hat{E}_i,
\label{eq:app-UD-POVM}
\end{eqnarray}
with sufficiently small $\lambda_i>0$ so that $E_\kappa \ge 0$.
Then, for $j\neq i$,
\begin{eqnarray}
\bra{\varphi_j}\hat{E}_i\ket{\varphi_j} = \lambda_i |\braket{\widetilde{\varphi}_i}{\varphi_j}|^2 = 0,
\end{eqnarray}
whereas, for $j=i$,
\begin{eqnarray}
\bra{\varphi_i}\hat{E}_i\ket{\varphi_i} = \lambda_i |\braket{\widetilde{\varphi}_i}{\varphi_i}|^2 = \lambda_i >0.
\end{eqnarray}
Hence, unambiguous discrimination exists.
\end{proof}

We now combine {\bf Lemma~\ref{lem:UD-linear-independence}} with a simple ``postprocessing'' argument.

\begin{lemma}[CP preprocessing cannot create unambiguous discrimination for dependent inputs]
\label{lem:cp-cannot-create-UD}
Let $\{\hat{P}_{\psi_j}\}_{j=1}^m$ be a set of pure states (rays). Suppose there exists a CPTNI map ${\cal E}$ such that: for each $j$,
\begin{eqnarray}
{\cal E}(\hat{P}_{\psi_j}) = p_j \hat{P}_{\varphi_j}, \quad p_j>0,
\label{eq:app-cp-map-pure-to-pure}
\end{eqnarray}
where $\{ \ket{\varphi_j} \}_{j=1}^m$ is a linearly independent set. Then, $\{ \ket{\psi_j} \}_{j=1}^m$ must also be linearly independent.
\end{lemma}

\begin{proof}---If $\{\ket{\varphi_j}\}$ are linearly independent, then by {\bf Lemma~\ref{lem:UD-linear-independence}} there exists a POVM $\{\hat{E}_1,\ldots, \hat{E}_m, \hat{E}_\kappa\}$ that unambiguously discriminates them.
Consider the following protocol for identifying an unknown input drawn from $\{P_{\psi_j}\}$: apply ${\cal E}$, and conditioned on success, perform the POVM $\{\hat{E}_i\}$ on the output. For input $\hat{P}_{\psi_j}$, the probability of returning outcome $i\neq j$ is
\begin{eqnarray}
\tr{\bigl(\hat{E}_i\,{\cal E}(\hat{P}_{\psi_j})\bigr)} = p_j \tr{\bigl(\hat{E}_i \hat{P}_{\varphi_j}\bigr)} = p_j \bra{\varphi_j}\hat{E}_i\ket{\varphi_j} = 0,
\end{eqnarray}
while the probability of correctly returning $j$ is
\begin{eqnarray}
\tr{\bigl(\hat{E}_j\,{\cal E}(\hat{P}_{\psi_j})\bigr)} = p_j \bra{\varphi_j}\hat{E}_j\ket{\varphi_j} > 0.
\end{eqnarray}
Hence, the composite ``CPTNI map + POVM'' protocol unambiguously discriminates the original set $\{\hat{P}_{\psi_j}\}$ with nonzero success probability. By {\bf Lemma~\ref{lem:UD-linear-independence}}, this is only possible if $\{\ket{\psi_j}\}$ are linearly independent.
\end{proof}

{\bf Lemma~\ref{lem:cp-cannot-create-UD}} is the abstract reason why the existence of a universal superposition machine would have immediate no-go consequences: if such a machine could take a linearly dependent family of inputs and output a linearly independent family with nonzero probability, it would contradict {\bf Lemma~\ref{lem:cp-cannot-create-UD}}. We now show that this is exactly what happens unless an ``overlap-determinability'' resource removes an otherwise unavoidable gauge freedom.

\subsection{Proof of Theorem~\ref{thm:overlap-resource}}\label{app:proof-overlap-resource}

We now prove {\bf Theorem~\ref{thm:overlap-resource}}. The central claim is that a nontrivial superposition map $\Lambda_{\alpha,\beta}$ cannot act on an overlap-indeterminate domain without forcing the forbidden transformation described in {\bf Lemma~\ref{lem:cp-cannot-create-UD}}. Our proof follows (and slightly reframes) the linear-independence argument of Ref.~\cite{Bandyopadhyay2020}, emphasizing the phase-gauge obstruction as the missing resource.

\vspace{2mm}
\noindent
{\bf Step 1: Choose a linearly dependent triple and an orthogonal ``phase-free'' partner.}
Assume $\dim{\cal H} \ge 3$~\footnote{The linear-independence reduction uses an auxiliary state $\ket{\phi}$ orthogonal to $\text{span}\{\ket{\psi}, \ket{\psi^\perp}\}$, hence we present it for $\dim{\cal H} \ge 3$. The case $\dim{\cal H} = 2$ follows from the case $\dim{\cal H} \ge 3$. Indeed, if a nonzero superposition map existed on a $2$-dimensional ${\cal H}$, then by embedding into a $3$-dimensional extension $\tilde{\cal H} = {\cal H} \oplus \mathbb{C}$ and letting the map act only on the embedded subspace (and fail otherwise), we would obtain a corresponding nonzero superposition map on $\tilde{\cal H}$, contradicting the argument here for $\dim\tilde{\cal H} \ge 3$.}. Choose two orthonormal vectors $\ket{\psi}$ and $\ket{\psi_\perp}$, and choose a third unit vector $\ket{\phi}$ orthogonal to both:
\begin{eqnarray}
\braket{\psi}{\psi_\perp}=0, \quad \braket{\phi}{\psi}=0, \quad \braket{\phi}{\psi_\perp}=0.
\label{eq:app-orthogonal-triple}
\end{eqnarray}
Now pick a third state in the span of $\ket{\psi}, \ket{\psi_\perp}$:
\begin{eqnarray}
\ket{\psi_3} = a\ket{\psi} + b\ket{\psi_\perp},
\label{eq:app-psi3}
\end{eqnarray}
where $a,b \neq 0$ and $\abs{a}^2 + \abs{b}^2=1$. The set $\{\ket{\psi},\ket{\psi_\perp},\ket{\psi_3}\}$ is linearly dependent by construction, since it lives in a two-dimensional subspace.

\vspace{2mm}
\noindent
{\bf Step 2: Apply the hypothetical superposition machine.}
Assume (towards contradiction) that {\bf Theorem~\ref{thm:overlap-resource}} fails, i.e., that there exists a superposition map $\Lambda_{\alpha,\beta}$ with $\alpha,\beta \neq 0$ that succeeds (with nonzero probability) on the three input pairs
\begin{eqnarray}
(\hat{P}_\psi,\hat{P}_\phi), \quad (\hat{P}_{\psi_\perp},\hat{P}_\phi), \quad (\hat{P}_{\psi_3},\hat{P}_\phi).
\label{eq:app-input-pairs}
\end{eqnarray}
Then, Eq.~(\ref{eq:app-superposer-def}) implies there exist phases $\theta_1,\theta_2,\theta_3$ such that the corresponding output vectors can be written as
\begin{eqnarray}
\ket{\Psi_1} &=& \alpha \ket{\psi} + \beta e^{i\theta_1}\ket{\phi}, \nonumber\\
\ket{\Psi_2} &=& \alpha \ket{\psi_\perp} + \beta e^{i\theta_2}\ket{\phi}, \nonumber\\
\ket{\Psi_3} &=& \alpha \ket{\psi_3} + \beta e^{i\theta_3}\ket{\phi} = \alpha(a\ket{\psi}+b\ket{\psi_\perp}) + \beta e^{i\theta_3}\ket{\phi}.
\label{eq:app-output-states}
\end{eqnarray}
Here, we intentionally keep $\theta_j$ as phases associated to the rays in Eq.~(\ref{eq:app-input-pairs}), consistent with the covariance requirement explained around Eq.~(\ref{eq:app-gauge}).

\vspace{2mm}
\noindent
{\bf Step 3: Show the outputs are necessarily linearly independent.}
We claim that $\{\ket{\Psi_1},\ket{\Psi_2},\ket{\Psi_3}\}$ are linearly independent for any such $\Lambda_{\alpha,\beta}$ when the domain is overlap-indeterminate, i.e., when the coefficients $a$ and $b$ in Eq.~(\ref{eq:app-psi3}) are not physical invariants available to the map. To see this, suppose the contrary: assume there exist complex numbers $x_1,x_2,x_3$, not all zero, such that
\begin{eqnarray}
x_1\ket{\Psi_1}+x_2\ket{\Psi_2}+x_3\ket{\Psi_3}=0.
\label{eq:app-lincomb}
\end{eqnarray}
Substituting Eq.~(\ref{eq:app-output-states}) into Eq.~(\ref{eq:app-lincomb}) and using the orthogonality in Eq.~(\ref{eq:app-orthogonal-triple}), we obtain
\begin{eqnarray}
\alpha(x_1+a x_3)\ket{\psi} +\alpha(x_2+b x_3)\ket{\psi_\perp} + \beta\left(e^{i\theta_1}x_1+e^{i\theta_2}x_2+e^{i\theta_3}x_3\right)\ket{\phi} = 0.
\label{eq:app-coeff-eq}
\end{eqnarray}
Since $\{\ket{\psi},\ket{\psi_\perp},\ket{\phi}\}$ are linearly independent, each coefficient must vanish:
\begin{eqnarray}
x_1 + a x_3 &=& 0, \label{eq:app-x1}\\
x_2 + b x_3 &=& 0, \label{eq:app-x2}\\
e^{i\theta_1}x_1+e^{i\theta_2}x_2+e^{i\theta_3}x_3 &=& 0. \label{eq:app-x3}
\end{eqnarray}
If $x_3=0$, then Eqs.~(\ref{eq:app-x1})--(\ref{eq:app-x2}) force $x_1=x_2=0$, contradicting that not all $x_j$ vanish.
Hence $x_3\neq 0$, and we may divide by $x_3$ and eliminate $x_1,x_2$ using Eqs.~(\ref{eq:app-x1})--(\ref{eq:app-x2}), yielding the necessary condition
\begin{eqnarray}
e^{i\theta_3} = a e^{i\theta_1} + b e^{i\theta_2}.
\label{eq:app-phase-condition}
\end{eqnarray}
This Eq.~(\ref{eq:app-phase-condition}) is where the ``missing resource'' appears. The coefficients $a$ and $b$ in Eq.~(\ref{eq:app-psi3}) are not physical invariants of the rays $(\hat{P}_\psi, \hat{P}_{\psi_\perp}, \hat{P}_{\psi_3})$ because the rays $\hat{P}_\psi$ and $\hat{P}_{\psi_\perp}$ admit independent phase gauges:
\begin{eqnarray}
\ket{\psi} \rightarrow e^{i\gamma_1}\ket{\psi}, \quad \ket{\psi_\perp} \rightarrow e^{i\gamma_2}\ket{\psi_\perp}.
\label{eq:app-independent-gauges}
\end{eqnarray}
These transformations do not change the input rays $\hat{P}_\psi$ and $\hat{P}_{\psi_\perp}$, but they change the coefficients in the decomposition of $\ket{\psi_3}$ as
\begin{eqnarray}
a \rightarrow a' = e^{-i\gamma_1} a, \quad b \rightarrow b' = e^{-i\gamma_2} b.
\label{eq:app-coeff-gauge-transform}
\end{eqnarray}
In an overlap-indeterminate setting, $\Lambda_{\alpha,\beta}$ cannot access $\gamma_1$ and $\gamma_2$ (they are not encoded in the rays), hence the phases $\theta_1,\theta_2,\theta_3$ in Eq.~(\ref{eq:app-output-states}) are fixed by the rays and cannot compensate for arbitrary changes in Eq.~(\ref{eq:app-coeff-gauge-transform}).

Then, let us observe that the linear dependence condition Eq.~(\ref{eq:app-phase-condition}) must be a physically meaningful statement about the rays $\{\hat{P}_{\Psi_1}, \hat{P}_{\Psi_2}, \hat{P}_{\Psi_3}\}$. Therefore, it cannot depend on the arbitrary gauge choice in Eq.~(\ref{eq:app-independent-gauges}). But, under the gauge change Eq.~(\ref{eq:app-independent-gauges}), the condition Eq.~(\ref{eq:app-phase-condition}) would become
\begin{eqnarray}
e^{i\theta_3} = a' e^{i\theta_1} + b' e^{i\theta_2} = e^{-i\gamma_1}a e^{i\theta_1} + e^{-i\gamma_2} b e^{i\theta_2}.
\label{eq:app-phase-condition-gauged}
\end{eqnarray}
Since $\gamma_1$ and $\gamma_2$ can be chosen independently and arbitrarily, Eq.~(\ref{eq:app-phase-condition-gauged}) cannot hold for all gauges unless one of $a$ or $b$ is zero. However, by construction in Eq.~(\ref{eq:app-psi3}), we chose $a,b \neq 0$. Hence, Eq.~(\ref{eq:app-phase-condition}) cannot be a gauge-invariant consequence of the physical action of $\Lambda_{\alpha,\beta}$. The only remaining possibility is that our assumption of linear dependence was false. Therefore, $\{\ket{\Psi_1},\ket{\Psi_2},\ket{\Psi_3}\}$ must be linearly independent.

\vspace{2mm}
\noindent
{\bf Step 4: Contradiction via Lemma~\ref{lem:cp-cannot-create-UD}.}
The map $\Lambda_{\alpha,\beta}$, when restricted to the second input fixed at $\hat{P}_\phi$, defines a CPTNI map
\begin{eqnarray}
{\cal E}(\cdot) := \Lambda_{\alpha,\beta}\big( (\cdot)\otimes \hat{P}_\phi \big),
\label{eq:app-restricted-map}
\end{eqnarray}
which by assumption succeeds with nonzero probability on the three inputs $\hat{P}_\psi$, $\hat{P}_{\psi_\perp}$, $\hat{P}_{\psi_3}$ and outputs the three pure states $\hat{P}_{\Psi_1}, \hat{P}_{\Psi_2}, \hat{P}_{\Psi_3}$. We have just shown that the output vectors $\{\ket{\Psi_1},\ket{\Psi_2},\ket{\Psi_3}\}$ are linearly independent. Thus, by {\bf Lemma~\ref{lem:cp-cannot-create-UD}}, this would imply that the input vectors $\{\ket{\psi},\ket{\psi_\perp},\ket{\psi_3}\}$ are also linearly independent. But they are linearly dependent by construction, contradicting {\bf Lemma~\ref{lem:cp-cannot-create-UD}}.

This contradiction completes the proof of {\bf Theorem~\ref{thm:overlap-resource}}.
\qed

\vspace{2mm}
\noindent
Now we highlight the physical interpretation of our overlap-determinability as the missing resource. The most important thing is the gauge freedom in Eq.~(\ref{eq:app-independent-gauges}): without an additional physical structure that ties together the phases of two rays, the decomposition coefficients $(a,b)$ in Eq.~(\ref{eq:app-psi3}) are not physical data, and the superposition map cannot choose output phases $\theta_j$ consistently. A reference state $\ket{\chi}$ with nonzero overlaps $\braket{\chi}{\psi} \neq 0$ and $\braket{\chi}{\phi} \neq 0$ precisely supplies such a structure, because it induces a physically meaningful phase relation between rays. This is why the known probabilistic constructions in Refs.~\cite{Oszmaniec2016,Dogra2018,Li2017,Hu2016} require nonzero (and effectively controllable) overlaps with a fixed referential state: they inject the missing overlap-determinability resource, thereby removing the gauge obstruction exploited above.

\subsection{Remark: connection to known probabilistic constructions}\label{app:remark-constructive}

For completeness, we recall how the overlap resource resolves the gauge problem in a constructive way. Let $\ket{\chi}$ be a fixed reference state. Whenever $\braket{\chi}{\psi} \neq 0$ and $\braket{\chi}{\phi} \neq 0$, the gauge-invariant superposition ray
\begin{eqnarray}
\hat{P}_{\Psi_\chi(\psi,\phi)} \ \ \text{with} \ \  \ket{\Psi_\chi(\psi,\phi)} \propto \alpha\,\frac{\braket{\chi}{\phi}}{|\braket{\chi}{\phi}|}\ket{\psi} + \beta\,\frac{\braket{\chi}{\psi}}{|\braket{\chi}{\psi}|}\ket{\phi}
\label{eq:app-gauge-invariant-superposition}
\end{eqnarray}
is well-defined, i.e., independent of the independent rephasings in Eq.~(\ref{eq:app-gauge}). The probabilistic protocol of Ref.~\cite{Oszmaniec2016} (and its experimental implementations \cite{Li2017,Hu2016}) realizes Eq.~(\ref{eq:app-gauge-invariant-superposition}) by entangling the two inputs with a control system (via a controlled-SWAP/Fredkin-type operation) and then postselecting on a projection of one register onto $\ket{\chi}$. In our language, the overlap-determinability resource is the availability of a nonvanishing reference overlap that operationally locks the phase gauge and prevents the contradiction in Eq.~(\ref{eq:app-phase-condition-gauged}).

%


\end{document}